\documentclass{ws-ijmpe}
\usepackage[super,compress]{cite}
\usepackage{graphicx}
\usepackage{float}
\usepackage{multirow}
\usepackage{longtable}
\usepackage{tabularx}
\usepackage{booktabs}
\usepackage{xcolor}
\usepackage{soul}
\usepackage{ulem}

\newcommand\Tstrut{\rule{0pt}{2.6ex}}

\begin{document}

\markboth{W. Asous, M. Syamimi, M.-H. Koh, K.-S. Khoo}{Deformed magic numbers at $N=178$ and $Z=120, 124$ in superheavy region}

\catchline{}{}{}{}{}


\title{Deformed magic numbers at $N=$178 and $Z=$120, 124 \\ in the 112 $\leq N \leq $ 190 superheavy region \\ from Skyrme mean-field calculations}

\author{W. Asous}
\address{{Department of Applied Physics, 
	Faculty of Science and Technology, Universiti Kebangsaan Malaysia (UKM), 43600 Bangi, Selangor, Malaysia.}\\wissalasous@gmail.com}


\author{Mastura Syamimi Abdullah}
\address{Department of Physics, Faculty of Science, Universiti Teknologi Malaysia, 81310 Johor Bahru, Johor, Malaysia.}

\author{Meng-Hock Koh\footnote{kmhock@utm.my}}
\address{Department of Physics, Faculty of Science, Universiti Teknologi Malaysia, 81310 Johor Bahru, Johor, Malaysia.\\ UTM Centre for Industrial and Applied Mathematics, Universiti Teknologi Malaysia, 81310 Johor Bahru, Johor, Malaysia.\\kmhock@utm.my}

\author{Kok-Siong Khoo}
\address{Department of Applied Physics, 
	Faculty of Science and Technology, Universiti Kebangsaan Malaysia (UKM), 43600 Bangi, Selangor, Malaysia. \\ Nuclear Technology Research Centre (NUKLEAR), Faculty of Science and Technology, Universiti Kebangsaan Malaysia 43600 Bangi. Selangor Malaysia.}

\maketitle

\begin{history}
\received{Day Month Year}
\revised{Day Month Year}
\end{history}

\begin{abstract}
\noindent{\textbf{Background:}} Various motivations for exploration of superheavy region revolve around the question 
on whether 126 is a spherical proton magic number, as is the case for neutrons.
In exploring this region, identification of nuclei with relatively longer half-life as compared to its neighbours
is crucial for experimental studies.
Such information are provided from theoretical predictions, which are however, heavily dependent on the theoretical model used
and observable quantities under investigation. \\
\noindent{\textbf{Purpose:}} {Limiting ourselves to the Skyrme Hartree-Fock-plus-Bardeen–Cooper–Schrieffer approach, we aimed to analyse the appearance of a nuclear region with relatively high stability
associated with emergence of spherical and deformed magic numbers in the region of $170 \leq N \leq 190$ ($112 \leq Z \leq 130$) based on various observables.} \\
\noindent{\textbf{Methods:}} Three Skyrme parametrizations namely the SkM* frequently employed for fission calculations, 
and the SLy5 and SLy4 commonly used for superheavy region, are considered to provide comparisons within the Skyrme mean-field approach.
We evaluated the variation of electric quadrupole deformation ($\beta_{20}$), binding energy per nucleon ($BE/A$), two-nucleon separation energy differential ($\delta S2_{q}$), 
alpha-decay energy ($Q_{\alpha}$) and alpha-decay half-lives ($T_{1/2}$). \\
\noindent{\textbf{Conclusion:}}  Our analyses suggest that neutron number $N = 178$ is candidate for deformed magic number 
around proton number $114 \le Z \le 118$.
For protons, $Z = 120$ and $124$ appears to be a candidate for deformed magic number at around $N = 172 \sim 178$.
Both sets of deformed magic numbers appear at oblate ground-state deformation.
\end{abstract}


\ccode{PACS numbers:}


\section{Introduction}\label{intro}

Investigation of superheavy (SH) region with atomic number of $Z \geq 104$ is fuelled by
the motivation to explore the limits of nuclear chart \cite{giuliani2019colloquium} at the upper right region.
The heaviest element produced thus far is the Oganesson with atomic number $(Z) = 118$ {\cite{ts2006synthesis, ts2012production, oganessian2015superheavy}}.
The basis for exploring this superheavy region is that the $126$ is expected to be magic for protons
as was the case for neutrons \cite{oganessian2015superheavy}.
In practice, one would require making multiple \emph{stops} before the $Z = 126$ can be synthesized.
These \emph{stops} refers to nuclei with specific combination of proton $Z$ and neutron $N$ numbers
which have sufficiently long half-life as compared to its neighbouring nuclei.
In this context, theoretical calculations to identify stable isotopes as compared to their neighbours 
would be handy as experimentalists explores the SH region.

However, theoretical predictions of nuclei with magic or sub-magic numbers 
are heavily model-dependent and often provide a conflicting results.
On the one hand, there are the microscopic calculations based on the mean-field and beyond mean-field frameworks
{\cite{rutz1997superheavy,gupta1997structure,bender1999shell,patra1999systematic,kruppa2000shell,berger2001superheavy,bhuyan2012magic,brodzinski2013predictions,li2014superheavy,shi2019superheavy,stone2019physics,biswal2020nuclear,sarriguren2021self,zhang2021predictive,jain2022exploring,sarriguren2022competition,dumitrescu2023cluster}}.
On the other hand, there are the microscopic--macroscopic models for e.g. those based on the cluster decay {\cite{xu2004alpha,xu2006new,ni2009microscopic,ni2009exotic,ni2010systematic,ismail2016relative,ismail2017alpha,ismail2020correlation}}, and
generalized- and effective- liquid drop model {\cite{wang2011alpha,wang2013alpha,cui2018alpha}}. Alpha-decay models such as the 
	Coulomb and proximity potential model and deformed nuclei (CPPM and CPPMDN) have also been used to explore the SHE region {\cite{santhosh2014probable,santhosh2014predictions,manjunatha2016comparison,santhosh2016feasibility,manjunatha2016alpha,manjunatha2016theoretical,santhosh2016theoretical1,santhosh2016theoretical2,santhosh2017decay,santhosh2018predictions1,santhosh2018predictions2,prathapan2021studies}}.

In all these studies, observable quantities of interest are indeed the 
alpha-decay namely $Q_{\alpha}$-values and half-lives $T_{1/2}$ since nuclei in this region tends to decay via alpha emission.
In addition, observables such as binding energy and
two-nucleon separation differential (shell-gap) have also been considered for e.g. in the work of Refs. {\cite{rutz1997superheavy,kruppa2000shell}}. 
Table~\ref{tab:list of magic candidates} shows a non-exhaustive list of proton and neutron
magic or sub-magic numbers obtained from these predictions.

For protons, we find that $Z = 114$ and $120$ are often reproduced by most methods.
In the case of neutrons, the predicted magic or sub-magic numbers are less certain
with estimated range between $N = 172$ to $N = 178$.
In addition to these sets of proton and neutron magic numbers, one should
also include the spherical doubly magic nucleus with $Z = 126$ and $N = 184$.
		
It is the purpose of this work to revisit the investigation of magic and sub-magic numbers
through cross-analyses of quantities which are sensitive to shell effect namely 
two-nucleon separation energies and single-particle levels, and, 
bulk properties in the form of alpha-decay energy $Q_{\alpha}$ and half-life $T_{1/2}$.
We have limit ourselves herein to nuclei in the region of $112 \leq Z \leq 130$ and $170 \leq N \leq 190$
which are expected to be relevant to the current experimental progress.

Our work relies on theSkyrme--Hartree--Fock--plus--Bardeen--Cooper--Schrieffer (SHF+BCS) framework which will be discussed in Section~\ref{mean-field}.
Calculations of $Q_{\alpha}$ and half-lives $T_{1/2}$ using the binding energies obtained from the SHF+BCS are then discussed in Section~\ref{formulae}.
Section~\ref{results} will be dedicated to the analyses 
of ground-state deformation, separation energies differential, single-particle levels, and alpha-decay properties. 
Finally, conclusion of the work will be discusses in Section~\ref{conclusion}.

\begin{table}[pt]
	\tbl{\label{tab:list of magic candidates} 
		Summary predictions of magic and sub-magic numbers through analyses of observable quantities based on the theoretical models of
		Hartree-Fock-plus-Bardeen–Cooper–Schrieffer (HF+BCS), Hartree-Fock-Bogoliubov (HFB), 
		Effective Liquid-Drop-Model (ELDM), Generalized-Liquid-Drop-Model (GLDM), 
		Weizsäcker–Skyrme 4 (WS4), Relativistic-Mean-Field (RMF), 
		Density-Dependent Cluster-Model (DD-CM), Macroscopic-Microscopic (Mac-mic), 
		Coulomb and Proximity Potential Model (CPPM) and for Deformed Nuclei (CPPMDN).}
		{\begin{tabular}{@{}ccccc@{}}
			\toprule 
		Model &  Quantity & Z  & N&  Refs.   \\
		\colrule \Tstrut
		\multirow{6}{*}{HF+BCS} &  $\delta_{2q}$ &  114, 126 &    184  &  
		{\cite{rutz1997superheavy}} \\ 
        & SPS & 114, 126 & 184 &  \cite{bender1999shell} \\
        & SPS & 120 & 172 & \cite{bender1999shell} \\
		&   SC & 124, 126 & 184   & {\cite{kruppa2000shell}} \\
		&  $\Delta_q$, $S_{2n}$, SPS    &    120    & 172, 182, 208  & {\cite{bhuyan2012magic}} \\
		&   SPS & 164 & 308  & {\cite{brodzinski2013predictions}}\\
		\colrule
		\multirow{2}{*}{HFB} & SPS  &  120 &  172  & {\cite{decharge2003bubbles}} \\ 
		& Shell-gap &  120, 126 &  172, 184  & {\cite{decharge1999superheavy,berger2001superheavy}}  \\
		\colrule
		ELDM &    $Q_\alpha$, $T_{1/2}$     &    120    &   178, 184 & {\cite{cui2018alpha}}  \\
		\multirow{2}{*} {GLDM}           &    $T_{1/2}$ &    114   &      184     & {\cite{wang2013alpha}} \\
		&    $T_{1/2}$ &  120, 126  &    184     & {\cite{wang2013alpha}} \\  
		\multirow{4}{*}{WS4} & {$Q_\alpha$, $T_{1/2}$} &   & 162, 178, 184 & {\cite{wang2015systematic}} \\
		& \multirow{2}{*} {$\delta_{2q}$} & 114  & 184 & {\cite{mo2014systematic}} \\ 
		&   & 120 & 178 & {\cite{mo2014systematic}} \\
		\colrule
		\multirow{9}{*}{RMF} &  $\delta_{2q}$, SPS &   120 & 172 & {\cite{rutz1997superheavy,bender1999shell}} \\
		&     SC        & 120  & 172  & {\cite{kruppa2000shell}} \\
		&   SPS & 120 &  172, 184 & {\cite{gupta1997structure}} \\
		&   $BE/A$, SPS & 114  &  164 $\sim$ 172 &  {\cite{patra1999systematic}} \\
		&   $BE/A$, SPS & 120 &  172, 184 & {\cite{patra1999systematic}} \\
		&   $\delta_{2n}$, SPS & -- &  168, 174, 178 & {\cite{siddiqui2021microscopic}} \\
		& \multirow{2}{*}{$\delta_{2q}$, SPS}  & \multirow{2}{*} {120, 138} & 172, 184,  & \multirow{2}{*}{\cite{li2014superheavy}} \\
		&      & & 228, 258    &        \\
		& $\Delta_q$, $S_{2n}$, SPS     & 120 &  172, 184, 258 &  {\cite{bhuyan2012magic}}      \\
		\colrule
		\multirow{8}{*} {DD-CM} & \multirow{7}{*}{$T_{1/2}$} & \multirow{4}{*}{} 82, 98 & \multirow{7}{*}{} 126, 148 &  \multirow{7}{*}{\cite{ismail2016relative}}\\
		&                       & 100, 102 & 152, 154 & \\
		&                       & 106, 108 & 160. 162 & \\
		&                       & 114, 116 & 172, 176 & \\
		&                       &          & 178, 180 & \\
		&                       &          & 180, 182 & \\
		&                       &          & 184, 200 & \\ 
		&    $T_{1/2}$     &    --      & 178  & {\cite{liu2019predictions}}  \\
		\colrule
		Mac-Mic & SC & 114, 120 & 174, 184 & \cite{malov2021shaping,malov2021landscape} \\  \colrule
		
		CPPMDN & $T_{1/2}$ & 120 & 178 & {\cite{santhosh2014predictions}} \\
		\multirow{2}{*}{CPPM} & $T_{1/2}$ & -- & 184 & {\cite{santhosh2014probable}} \\
		& $T_{1/2}$ & -- & 184, 202 & {\cite{santhosh2017decay}} \\
			\botrule
		\end{tabular}}
\end{table}

\section{Theoretical Framework}
\subsection{The Skyrme mean-field model}
\label{mean-field}

The standard Skyrme HF+BCS model is used in our study. 
We considered primarily
two Skyrme parametrizations namely the SkM* {\cite{skyrme1958effective} and the SLy5 \cite{chabanat1998skyrme}}.
The SkM* parametrization was fitted in the early 1980s and has been commonly employed in nuclear fission study.
As was the case for other parametrizations fitted in the early years of nuclear mean-field calculations,
the spin-orbit density $J$ term was neglected in the SkM* fit procedure.
In constrast, the SLy5 parametrization fitted in the 1990s takes into account the spin-orbit density term.
Apart from this difference, the fit procedures for both parametrizations are similar in that they 
use a density dependence of $\alpha = 1/6$ and the Coulomb exchange term is treated using the Slater approximation.
It is also interesting to note that the SLy5 parametrization was found to suffer from spin-instability 
arising from the $\mathbf{s}\Delta \mathbf{s}$ term in the Skyrme energy density functional.
The SLy5 parametrization was then refitted by \cite{pastoresly5} to overcome this issue leading to the new SLy5* parametrization.

In addition, we also compared the results with those obtained using the SLy4 parametrization which was reported 
to perform well in the superheavy region {\cite{cwiok1999structure,peng2022potential}}.

For the particle-hole part, we employed the seniority force where 
the pairing matrix element is given by:
\begin{equation}
	v_{k\bar{k}l\bar{l}} = \frac{G_q}{11+Nq},
\end{equation}
with $G_q$ and $N_q$ being the pairing strength
and total nucleon number of a given charge state $q \equiv \{n,p\}$.

The pairing strength $G_q$ was initially fitted such that the BCS pairing gap
reproduces the empirical Madland formula of each nucleus given by:
\begin{equation}
	\Delta_n = r \: \mbox{exp}\Big(\frac{-sI-tI^2}{N^{1/3}}\Big),
	\label{eq: Madland neutron}
\end{equation}
\begin{equation}
	\Delta_p = r \: \mbox{exp}\Big(\frac{sI-tI^2}{Z^{1/3}}\Big).
	\label{eq: Madland proton}
\end{equation}
The constants entering the formula are $r = 5.72$ MeV, $s = 0.118$ and $t = 8.12$ while $I = \frac{N-Z}{A}$ is the neutron excess.

Fitting $G_q$ in such a manner may yield values 
which are artificially increased especially when 
the single-particle level density at the mean-field level is not correctly reproduced.
We therefore took an average pairing strength for the neutron and proton channels separately,
of the 86 superheavy nuclei considered herein.
The retained values (in MeV) are shown in Table~\ref{tab:pairing strengths}.
The pairing strengths for the SLy4 parametrizations are taken to be the same as those of the SLy5 values.

\begin{table}[pt]
\tbl{\label{tab:pairing strengths}Averaged pairing strengths of SkM* and SLy5 Skyrme parametrizations taken from 86 nuclei whereby the pairing strengths for each nucleus have been fitted to the Madland pairing gap formula.}
{\begin{tabular}{@{}ccc@{}} \toprule
Skyrme type & $G_n^{Avg.}$ (MeV) & $G_p^{Avg.}$ (MeV) \\
			\colrule
			SkM*	& 14.79 			 & 13.72   			  \\ 
			SLy5 		& 16.72 			 & 14.25    	\\
            SLy4    & 16.72 			 & 14.25    	\\ 
            \botrule
\end{tabular}}
\end{table}

The single-particle wave-function is expanded on the harmonic oscillator basis states 
and the expansion is truncated such that:
\begin{equation}
	\hbar \omega_{\bot} (n_{\bot} + 1) + \hbar \omega_z (n_z + \frac{1}{2}) \le \hbar \omega_0 (N_0 +2),
\end{equation}
with the basis size of $N_0 = 18$. The spherical angular frequency $\omega_0^3 = \omega_{\bot}^2 \omega_z$ is related to the angular frequency $\omega_{\bot}$ in the $x-y$ plane and and $\omega_z$  in the $z$-direction.

The basis size is selected through comparison on $Q_\alpha$ values as a function of $N_0$ in  Og-294 isotope. 
These values are tabulated in Table \ref{tab:basis size}. 
It can be seen that the $Q_{\alpha}$ remains almost constant after $N_0 = 16$ with the SkM* parametrization
and after $N_0 = 18$ with the SLy5 parametrization.
Therefore, we opt to retain $N_0 = 18$ for both forces.

\begin{table}[pt]
\tbl{\label{tab:basis size}Basis size test on Og-294 with SkM* and SLy5 Skyrme parametrization.}
{\begin{tabular}{@{}ccc@{}} \toprule
$N_0$ & $Q_{\alpha}^{SkM^*}$ (MeV)  & $Q_{\alpha}^{SLy5}$ (MeV) \\
			\colrule
			14   &  11.86        & 11.66    \\ 
			16   &  12.18        & 11.74    \\  
                18   &  12.17        & 11.40    \\			
                20   &  12.20        & 11.43    \\
			22   &  12.22        & 11.41    \\ \botrule
\end{tabular}}
\end{table}

The numerical integration is handled using Gauss-Hermite and Gauss-Laguerre approximations
with 50 and 20 points, respectively.
The exchange Coulomb term is calculated using the usual Slater approximation \cite{slater1951}.
The ground-state is assumed to be axially and
parity symmetric shape \cite{flocard1973nuclear}.

\subsection{Alpha-decay half-lives formulas}\label{formulae}
We use the absolute value of the binding energy ($E_B$) 
from the Skyrme mean-field model to calculate the
$Q_{\alpha}$ using the expression: 
\begin{equation}
	\label{eq:Q-alpha}
	Q_{\alpha}(Z,N)= E_B(2,2)+E_B(Z-2,N-2)-E_B(Z,N),
\end{equation}
where $E_B(2,2) = 28.3$ MeV is the binding energy of He-4 nucleus.
We consider only the case whereby the alpha decay occurs from the ground-state of parent nucleus
to the ground-state of daughter nucleus.
The value of $Q_{\alpha}$ enters the alpha decay half-life formulas as discussed in the following subsections.

\subsubsection{Viola-Seaborg (VS) formula}

The Viola-Seaborg formula \cite{viola1966nuclear} was developed based on the Gamow model
with the half-life given by the expression:
\begin{equation}
	\log_{10}{T_{1/2}^{VS}=\left(aZ+b\right)}Q^{-\frac{1}{2}}+cZ+d+h.
	\label{eq:VS}
\end{equation}
Over the years, the original parameters of \cite{viola1966nuclear} have been refitted as new experimental data emerged.
We considered herein two sets of parameters to the Viola-Seaborg formula.
The first is referred to as the VSS1 taken from 
the work of Sobiczewski and Cwiok {\cite{sobiczewski1989deformed}}. 
The parameters are $a = 1.66175$, $b = -8.5166$, $c = -0.20228$, and $d = -33.9069$.
Another set of parameters referred to as VSS2 herein 
are taken from {\cite{parkhomenko2005phenomenological}}
where the values are $a = 1.3892$, $b = 13.862$, $c = -0.1086$ and $d = -41.458$. The hindrance factor $h$ is set to 0 for even-even nuclei in both cases.

\subsubsection{Parkhomenko--Sobiczewski (PS) formula}Parkhomenko and Sobiczewski \cite{parkhomenko2005phenomenological,sobiczewski2007description} dropped the parameter $b$
from equation~\ref{eq:VS} and refitted the resulting half-life equation:
\begin{equation}
	\log_{10}{T_{1/2}^{PS}=aZQ_\alpha^{-\frac{1}{2}}+bZ+c},
\end{equation}
using the same 61 nuclei with $Z = $ 84 -- 110 and $N = $ 128 -- 160.
The retained values for the adjustable parameters are 
$a = 1.537$2, $b = -0.1607$ and $c = -36.573$.
The root-mean-square (r.m.s) was reported to differ only very slightly as compared to the
four parameters set of VSS2.

\subsubsection{Royer (R) and modified Royer (mR) formulas}
The Royer's formula \cite{royer2000alpha} developed based on the liquid drop model is given by the expression:
\begin{equation}
	\log_{10} {T_{1/2}^{R}}=aZQ^{-\frac{1}{2}}+bA^\frac{1}{6}Z^\frac{1}{2}+c.
	\label{eq:royer}
\end{equation}
We selected the values of adjustable parameters corresponding to the fit to 131 even-even alpha emitters \cite{royer2000alpha} 
with $a = 1.5864$, $b = -1.1629$, and $c =-25.31$.

We have also considered a modified Royer formula developed by \cite{royer2010analytic} 
where the angular momentum of the $\alpha$-particle is taken into account.
The modified formula takes the form of
\begin{flalign}
	 \log_{10}{T_{1/2}^{mR}}=aZQ^{-\frac{1}{2}}+bA^\frac{1}{6}Z^\frac{1}{2}+c+d\frac{\left(ANZ\left[l\left(l+1\right)\right]^\frac{1}{4}\right)}{Q}+eA[1-\left(-1\right)^l].
	\label{eq:modified royer}
\end{flalign}
For the case of ground-state--to--ground-state transition between the parent and daughter nuclei,
the angular momentum $l$ carried by the alpha particle is zero.
Therefore, the last two terms on the right-hand side of equation~(\ref{eq:modified royer}) vanishes in our case.
This gives rise to the same expression given in equation~(\ref{eq:royer}) with only a slight difference 
in the values of the adjustable parameters.
We employed the values of 
	$a = 1.58439$, $b = 1.15847$, $c = -25.31$ 
	fitted to 137 even-even nuclei in the work of {\cite{akrawy2018systematic}}.

\subsubsection{Brown and modified Brown formulas}

The Brown's formula \cite{brown1992simple} for alpha decay half-life is given by:
\begin{equation}
	\log_{10}{T_{1/2}^{BF}}=9.54\left(Z-2\right)^{0.6}Q^{-\frac{1}{2}}-51.37,
\end{equation}
whereby parameters were fitted to alpha emitters in the $Z = 76 - 108$ and $N = 90 - 156$ region.

We have also considered two modified Brown formula proposed by \cite{budaca2016extended}
fitted to 96 superheavy nuclei.
The first formula referred to as MB1 given by the expression: 
\begin{equation}
	\log{T_{1/2}^{MB1}}=a\left(Z-2\right)^bQ^{-\frac{1}{2}}+c+h,
\end{equation}
and was fitted by considering $a$, $b$ and $c$ to be the same for all sample nuclei 
while the hindrance factor $h$ to be dependent on whether the neutron and proton numbers
are even or odd. 
The parameters for MB1 are $a = 13.0705$, $b = 0.5182$ and $c = -47.8867$, while $h$ for even-even nuclei is set to 0.

The second modified Brown formula \cite{budaca2016extended} is given as:
\begin{equation}
	\log{T_{1/2}^{MB2}}=a\left(Z-2\right)^bQ^{-\frac{1}{2}}+c,
\end{equation} 
and was fitted without the hindrance factor but separated into four different sets corresponding to even-even, odd-even, even-odd and odd-odd nuclei. For the set of even-even nuclei, the parameters are
$a = 10.8238$, $b = 0.5966$, and $c = -56.9785$.

\subsection{Two-nucleon separation energy and separation energy differential}

The two-nucleon separation energy calculated using the equation (for the case of neutron):
\begin{equation}
	S_{2n} (N,Z) = E_{B}(N-2,Z)-E_{B}(N,Z),
	\label{eq: S2n}
\end{equation}
where $E(N,Z)$ is the binding energy of the nucleus with $Z$ protons and $N$ neutrons
is a frequent observable used to locate magic numbers.

The variation of the separation energy with consecutive nucleon number is another way in which one can determine magic numbers for e.g. as in the studies of {\cite{rutz1997superheavy, stone2019physics, kelvin2021role}}.
We refer to this quantity as the two-nucleon separation energy differential $\delta S_{2q}$.

As an example in the case of neutrons, the $\delta S2_{2n}$ is calculated using the equation:
\begin{equation}
	\delta S_{2n} = S_{2n}(N,Z)-S_{2n}(N+2,Z).
	\label{eq: delta S2n}
\end{equation}
The $\delta S_{2q}$ herein refers to the same quantity known as two-nucleon gap 
in \cite{rutz1997superheavy} given in terms of the three-point formula in the case of neutrons by:
\begin{flalign}
	\delta S_{2n} (N,Z) = 2 E_B(N,Z) - E_B(N-2, Z) - E_B(N+2,Z).
\end{flalign}

\section{Results and Discussions}\label{results}
\subsection{Location of the ground-state deformation $\beta_{20}$
\label{sec:results: ground state}}

We first performed multiple constrained calculations by fixing electric quadrupole moment ($Q_{20}$) while restricting only to axial and parity symmetric nuclear shapes.
The location of ground-state (g.s.) deformation is then obtained by releasing the constraint on $Q_{20}$ from the solution closest to where the minimum is expected.

\begin{figure}\centering
    \includegraphics[width=0.95\textwidth]{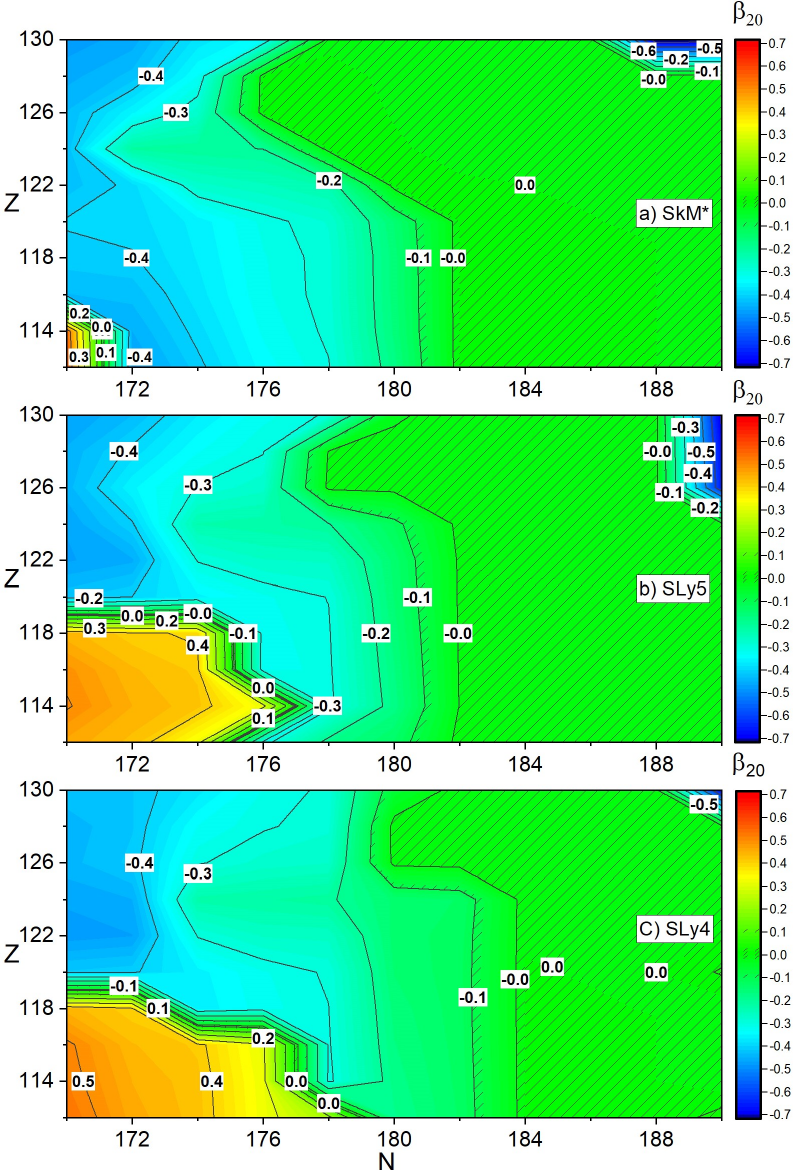}
    \caption{Contour map showing the location of the ground-state electric quadrupole moment (in dimensionless $\beta_{20}$) as a function of neutron $N$ and proton $Z$ numbers. Negative (positive) values indicate an oblate (prolate) shapes while $\beta_{20} = 0$ indicate a spherical shape.\label{fig: contour gs shape}}
\end{figure}

We plotted a contour map of the ground-state $\beta_{20}$ as a function of neutron number, $N$ in the range of $170 \le N \le 190$ and proton number, $Z$ in the range of $112 \le Z \le 130$ in Figure~\ref{fig: contour gs shape}. The dimensionless deformation parameter $\beta_{20}$ is derived from the quadrupole moment $Q_{20}$ given as {\cite{maruhn2005dipole}}:

\begin{equation}
    \beta_{20}=\frac{4 \pi}{5} \frac{Q_{20}}{A \langle r^2 \rangle} 
\end{equation}

\noindent where $ \langle r^2 \rangle$ is the root-mean-square radius $(rms)$, $\frac{4 \pi}{5}$ the scaling factor and $A$ the mass number.

All Skyrme parametrizations showed almost the same pattern in the evolution of the ground-state shape with a major distinction 
in the lower left region of the plot.
While all of the Skyrme parametrizations predict that nuclei located in the lower left region is prolate, 
the region with prolate deformation is significantly larger in the case of SLy5 and SLy4 (up to around $Z = 118$ and $N = 176$) as compared to SkM* (around $Z = 114$ and $N = 170$). 

Apart from that, another minor difference is in terms of the on-set of sphericity around $N = 184$.
For the SkM* and SLy5 parametrization, nuclei in the region around $114 \le Z \le 120$ are expected to be spherical starting at $N = 182$. However, sphericity is only achieved in the case of the SLy4 parametrization at $N = 184$ although the deformation at $N = 182$ is rather small. On the far right region at $Z = 130$ and $N = 190$ all of the considered Skyrme parametrizations 
predicted a rather rigid oblate deformation.

\subsection{Binding energy per nucleon at the ground-state deformation}
We tabulated the ground-state binding energy per nucleon ($\text{BE/A}$ in absolute value) obtained in our calculations
with those of Ref.{\cite{karim2019self}} using the same Skyrme parametrizations but within the 
Hartree-Fock-Bogoliubov (HFB) approach in Table~\ref{tab:binding energy per nucleon}.
In their calculations, the density-dependent delta pairing force was employed.

\begin{figure}[hbt!]\centering
	\includegraphics[width=1\textwidth]{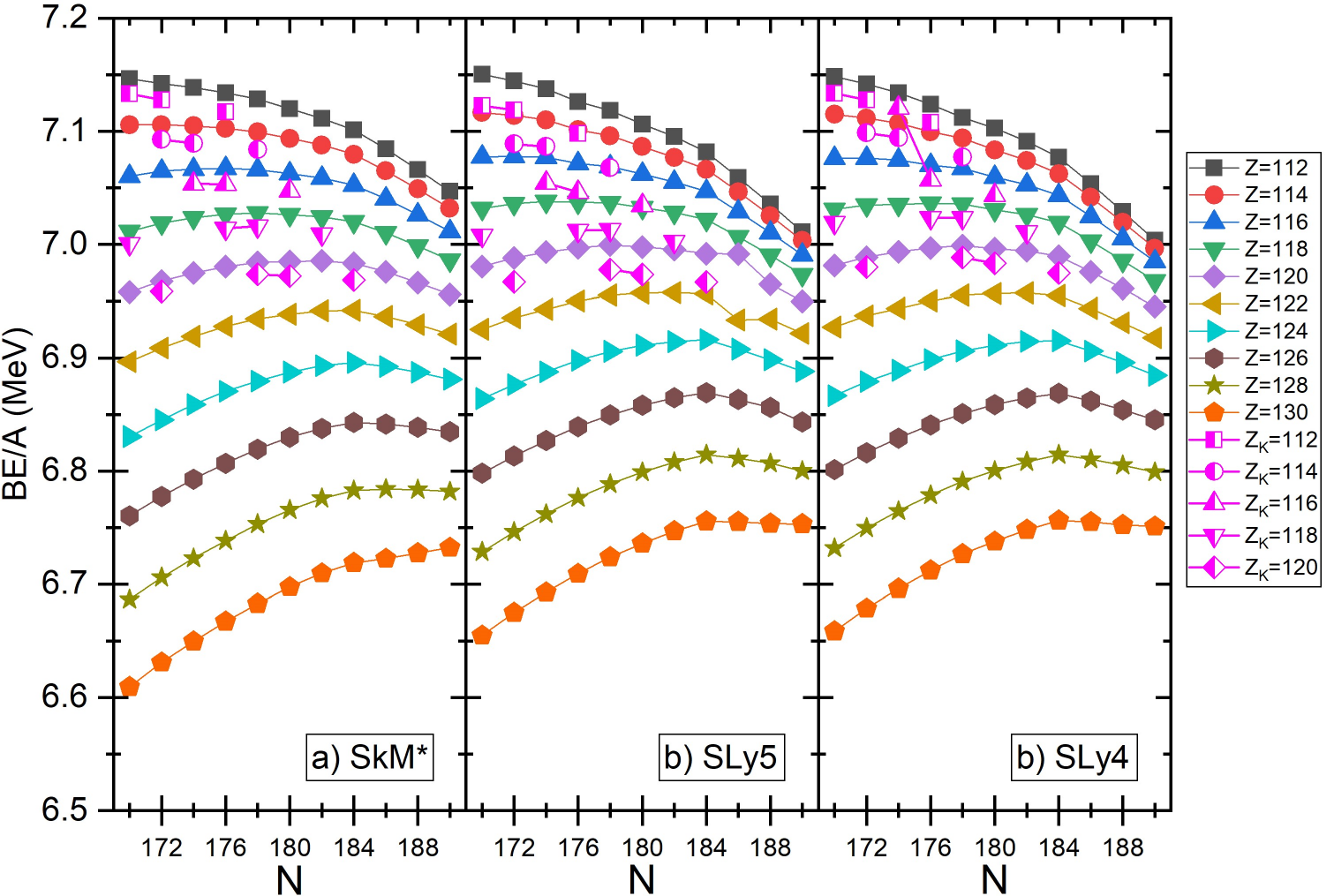}
	\caption{Shows plots of binding energy per nucleon ($BE/A$) for all Skyrme parametrization considered herein for $112 \leq Z \leq 130$. The obtained values are compared to the work of Ref. \cite{karim2019self} represented in half-filled magenta color.}
	\label{fig:binding energy per nucleon}
\end{figure}

We find that our results agree well with the HFB results where the differences in $\text{BE/A}$ are, on average, 
about 14 keV for SkM* and SLy4, and about 25 keV for SLy5.
Figure \ref{fig:binding energy per nucleon} shows the evolution of $\text{BE/A}$ as a function of mass number $A$.
In most cases, our calculated data shows the same trend as those obtained in the work of Ref.\cite{karim2019self}.
The only difference in the trend from one nucleus to another 
between our results and theirs appear between the $^{290}$Lv and $^{292}$Lv nuclei.
The HFB calculations showed a sudden drop in $\text{BE/A}$ with the SLy4.
This is not visible in our $\text{BE/A}$ results which varies smoothly with $A$.
Some tests have been performed, but the same pattern persists.

\subsection{Root-mean-square and charge radii}

\begin{figure}[hbt!]\centering
	\includegraphics[width=1\textwidth]{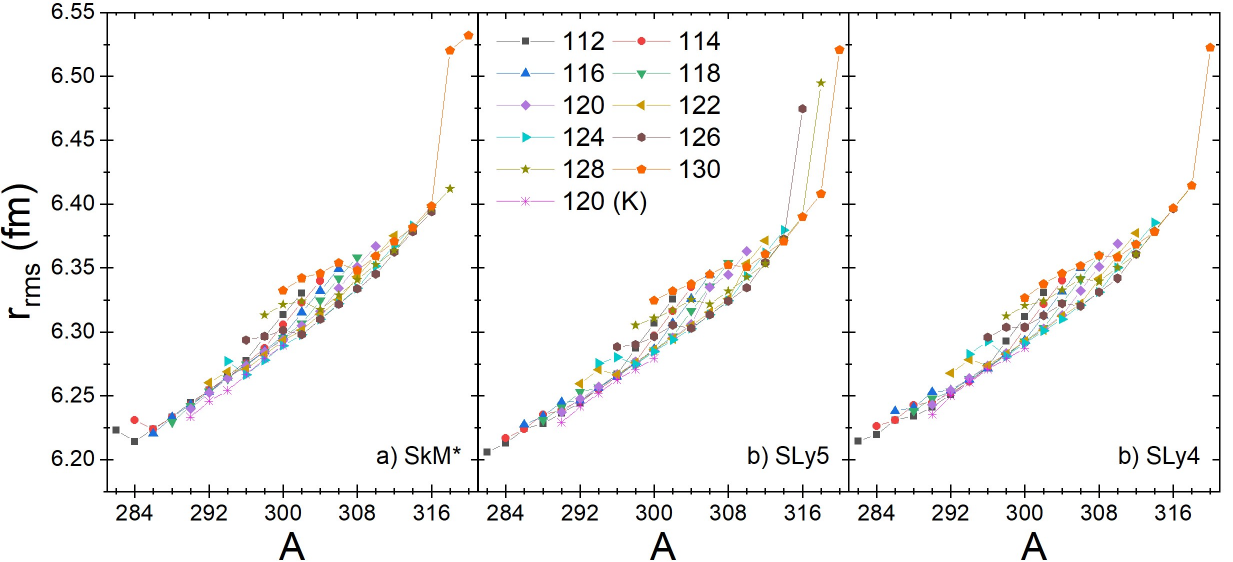}
    \includegraphics[width=1\textwidth]{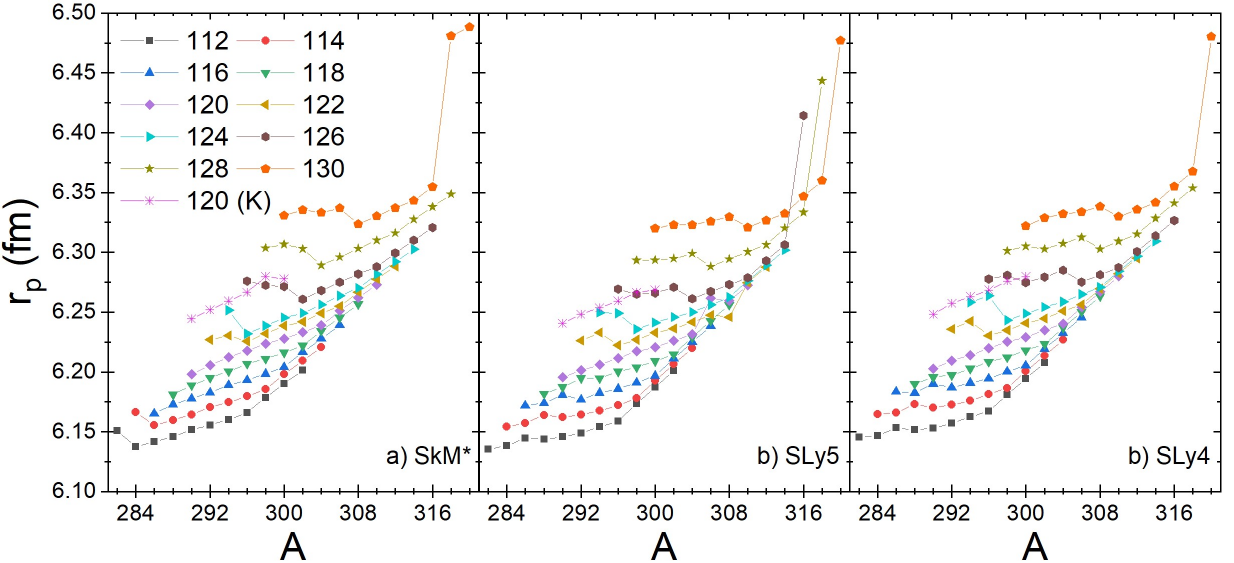}
	\caption{Top plots shows root-mean-square radii ($r_{rms}$) and bottom plots proton radius ($r_{p}$) for sample nuclei and Skyrme forces considered herein. The obtained values for $Z=120$ are compared to the work of Ref. \cite{karim2019self} represented in star-like marker in magenta color. }
	\label{fig:r_ch and r_rms all}
\end{figure}

We tabulated the root-mean-square radii ($r_{rms} = \sqrt{\langle \hat{r}^2 \rangle}$) 
and proton radius ($r_{p} = \langle \hat{r}_p^2 \rangle$) in 
Table \ref{tab: root-mean-square radii} and Table \ref{tab: charge radius} and plotted it in Fig.~\ref{fig:r_ch and r_rms all} for easy comparison.
We find that the $r_{rms}$ increases smoothly with $A$ as expected.
However, there appears a jump in $r_{rms}$ for nuclei at the end of the isotopic chain 
of $Z = 130$ (SkM*), $Z = 126, 128, 130$ (SLy5) and $Z = 128, 130$ (SLy4).
The change in $r_{rms}$ corresponds to the change in the ground-state deformation as seen
in Fig.~\ref{fig: contour gs shape} at the corresponding neutron and proton numbers.
Similiarly, the sudden jump in the case of $r_{p}$ occurs at the point where there is a change in the
ground-state deformation.

\begin{figure}[hbt!]\centering
	\includegraphics[width=0.8\textwidth]{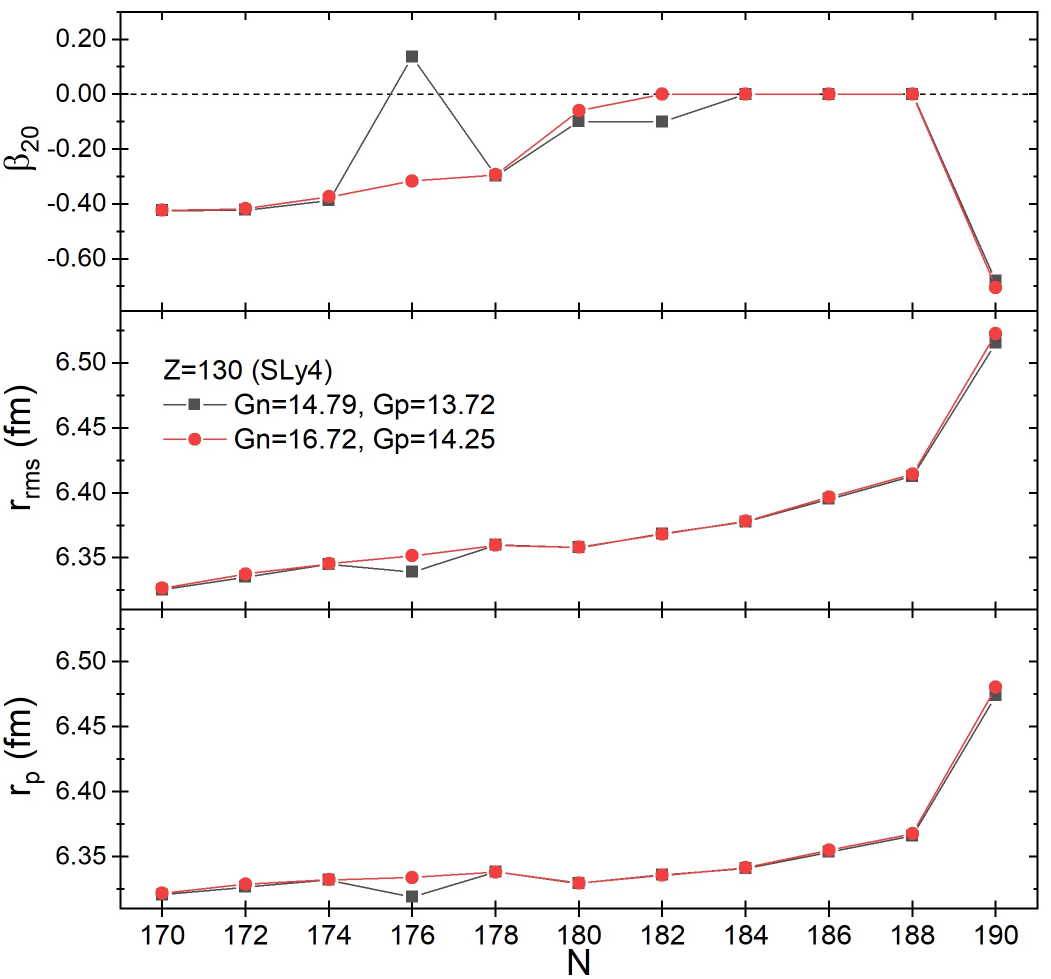}
	\caption{The figure shows plots for (a) root-mean-square radii $(r_{rms)}$ and (b) charge radii $r_{p}$ (fm) for SLy4 parametrization for $Z=130$. 
    The obtained values are compared to the work of Ref. \cite{karim2019self} represented in half-filled magenta color.}
	\label{fig:r_rms & r_ch pairing stregnth}
\end{figure}

It should be noted that the increase in $r_{rms}$ and $r_{p}$
is strongly affected by the choice of pairing strengths.
Fig.~\ref{fig:r_rms & r_ch pairing stregnth} shows a comparison of $r_{rms}$ (middle panel) and $r_{p}$ (bottom panel)
for $Z = 130$ isotopes obtained with two different sets of pairing strengths.
When $(G_n,G_p) = (16.72, 14.25)$ MeV are chosen, we find a jump in both $r_{rms}$ and $r_{p}$
at $N = 190$ (corresponding to $A = 320$ in Fig.~\ref{fig:r_ch and r_rms all}).
However, the $r_{rms}$ and $r_{p}$ varies smoothly with $N$ when lower values of $(G_n,G_p) = (14.79, 13.72)$ MeV
are employed.
This reiterates the importance of proper selection of pairing strengths.
Further study along the work of one co-author~\cite{2024_Koh} could be explored in future work.

In the earlier work\cite{karim2019self}, similar calculations for $Z = 120$ isotopes have been performed
where the charge radius $r_{ch}$ had been calculated using $r_{ch} = \sqrt{r_p^2 + 0.64}$.
These values obtained within the Hartree-Fock-Bogoliubov (HFB) framework are plotted in Fig.~\ref{fig:r_ch and r_rms all}
for comparison.
Apart from the constant $0.8^2 = 0.64$ to account for the finite size of the proton,
we find that the trend for the $Z = 120$ isotopes from the HFB calculations are similar to our results.

\subsection{Two-nucleon separation energy differential $\delta S_{2q}$} \label{sec:separation energy}

\begin{figure}[hbt!]\centering
	\includegraphics[width=1.0655\textwidth]{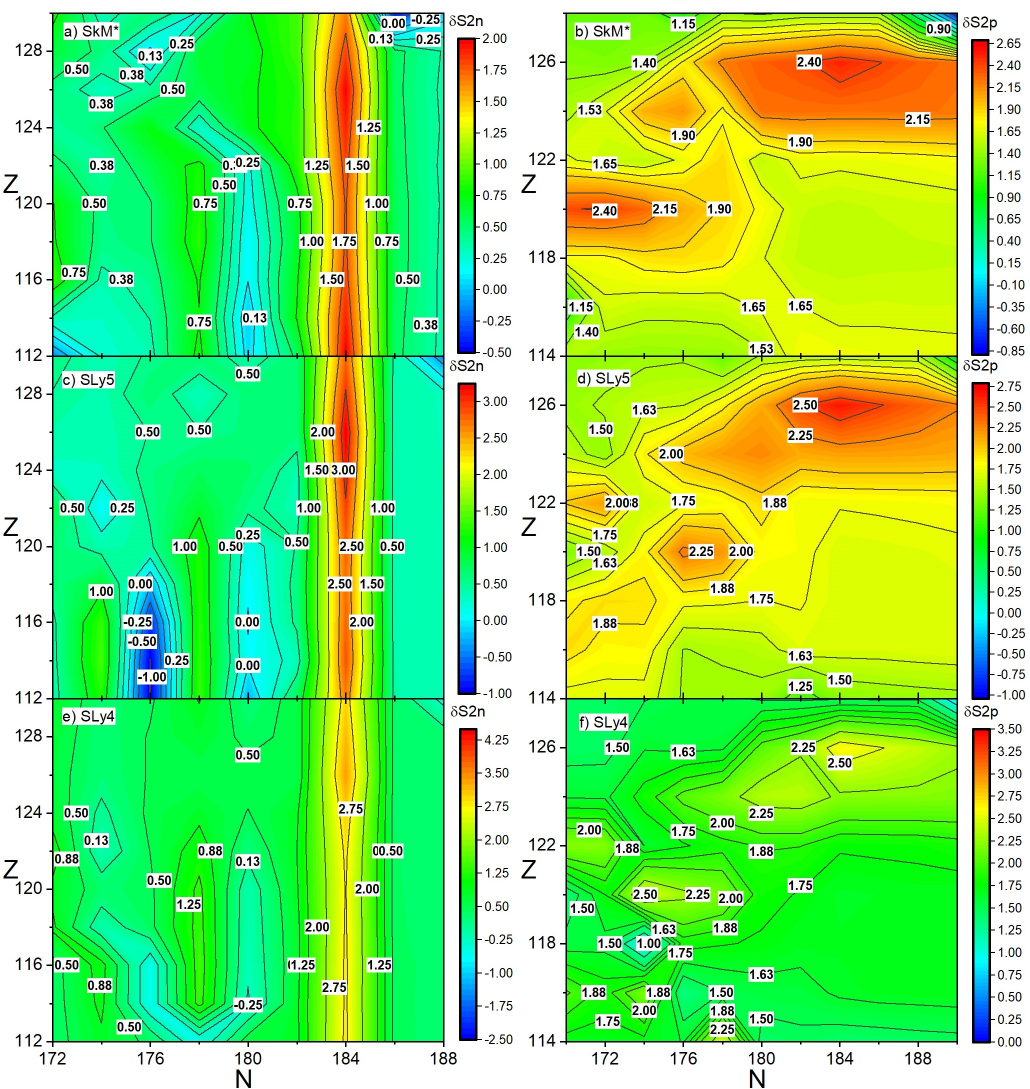}
	\caption{Shows contour mapping of $\delta S_{2n}$ left panel and $\delta S_{2p}$ panel. Skryme  parametrizations calculated shown at different rows (top) for SkM*, 
    (middle) SLy5 and (bottom) SLy4.}
	\label{fig:contour s2n s2p}
\end{figure}

Figure~\ref{fig:contour s2n s2p} shows the two-neutron separation energy differential  $\delta S_{2n}$ 
and two-proton separation energy differential $\delta S_{2p}$.
Simultaneous evaluation of both quantities is necessary to pinpoint the emergence of shell closures, as suggested by {\cite{rutz1997superheavy}}.
It is obvious from Figure~\ref{fig:contour s2n s2p} that all three Skyrme forces predict $Z = 126$ and $N = 184$ 
to be the next spherical magic numbers after $Z = 82$ and $N = 126$, this has indeed been proposed by many for e.g. {\cite{cwiok1996shell, rutz1997superheavy, cwiok2005shape}}.

Beyond this set of spherical magic numbers, there also appears regions with relatively high $\delta S_{2q}$.
Comparing the contour of $\delta S_{2n}$ (left panels) of
Figure~\ref{fig:contour s2n s2p}, 
there appears a ridge at $N = 178$ from $Z = 112$ to $122$ for all Skyrme forces.
In addition to $N = 178$, the SLy5 parametrization also predicts a region with high $\delta S_{2n}$ at $N = 174$
starting from $Z = 112$ to $118$.
The SLy4, however, shows only a small region of high $\delta S_{2n}$ at $N = 174$ for $Z = 114$ and $116$.
The region of high $\delta S_{2n}$ then shifted towards $N = 172$ and moving upwards from $Z = 118$ towards $Z = 130$.

For protons, the $\delta S_{2p}$ of Figure~\ref{fig:contour s2n s2p} (right panels)
shows a region with high $\delta S_{2p}$ comparable to the spherical magic number 
of $Z = 126$ at $Z = 120$ for $N = 172$ (SkM*), $N = 176$ (SLy5), and
$N = 174$ (SLy4).
For the SLy4, an even intense peak is seen at around
$Z = 114$, $N = 174$ suggesting a potentially relatively stable isotope due to proton magicity.
This is however not seen with the other two Skyrme forces.

Evaluating Figure~\ref{fig:contour s2n s2p} with respect to the $Q_{20}$ plot in Figure~\ref{fig: contour gs shape}, 
we identify $N = 178$ as a \textit{deformed magic number} from $Z = 114$ to $120$ with an oblate ground-state shape.
The emergence of $N = 174$ as deformed magic number 
predicted with the SLy5 and SLy4 appears at the prolate side of the deformation. 
Similarly, the intense peak at $Z = 114$, $N = 174$ predicted only with the SLy4
has the same deformation type.
On the other hand, the region with high $\delta S_{2n}$ at $N = 172$ from $Z = 118$ to $Z = 130$ 
predicted with the SLy4
has a prolate shape at $Z = 118$ which then shifted to an oblate shape starting at $Z = 120$.

For protons, the SkM* parametrization predicted a deformed magic number at $Z = 120$ around $N = 172$ with a rather strong oblate ground-state shape, 
and $Z = 124$ around $N = 176$ with a slightly oblate ground-state shape. 
The SLy5 parametrization predicted the location of deformed magic number at the same proton number but with different neutron number 
namely at $Z = 120$ around $N = 176$, 
and, $Z = 124$ around $N = 180$ with similar ground-state deformations as obtained with the SkM* parametrization.
The SLy4 predicts deformed proton magic number at $Z = 114$ around $N = 174$ with a prolate shape,
while $Z = 120$ with $N = 174$
and at $Z = 122$ with $N = 172$ with an oblate shape.

\subsection{Single-particle levels
	\label{sec: sp levels}}

\begin{figure}[hbt!]\centering
	\includegraphics[width=0.95\textwidth]
    {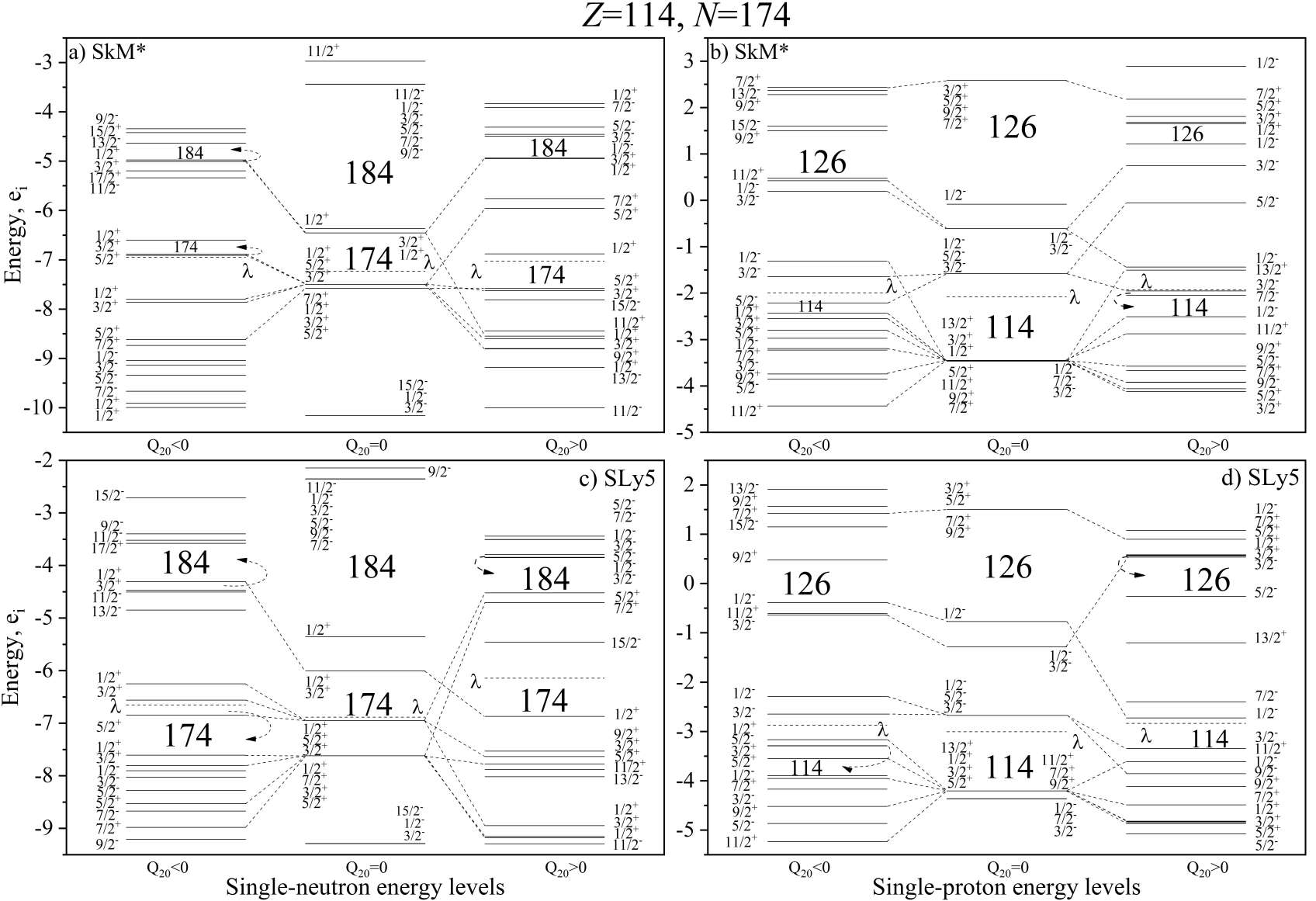}
    \includegraphics[width= 0.5\textwidth]{Figure_7_124_180_proton_sp_levels}
	\caption{Single-particle levels at the oblate ($Q_{20} < 0$), spherical ($Q_{20} = 0$) and prolate ($Q_{20} > 0$) minimum 
		obtained with the SkM* and SLy5 parametrizations for the $Z = 114$, $N = 174$ nucleus. The single bottom panel shows proton single-particle levels obtained with the SLy5 parametrization
		for the $Z = 124$ and $N = 180$ nucleus showing an energy gap at $Z = 124$ which does not appear in Figure~\ref{fig: sp levels Z = 114 N = 174} for the case of $N = 174$.}
	\label{fig: sp levels Z = 114 N = 174}
	\label{fig: protons sp levels of Z = 124 N = 180}
\end{figure}

Figure~\ref{fig: sp levels Z = 114 N = 174} shows comparison of the single-neutron and single-proton levels 
at the oblate, spherical and prolate shapes for $_{114}^{288}\mbox{Fl}_{174}$ nucleus 
obtained with the SkM* and SLy5 parametrizations. 
Here we show that the appearance of  ridges in the $\delta S_{2n}$ and $\delta S_{2p}$ plots 
is attributed to a substantial energy gap in the single-particle levels at the relevant nucleon numbers.

For neutrons, the SkM* parametrization shows that energy gaps for $N = 174$ at the prolate deformation 
while the SLy5 parametrization shows a substantial gap at the same neutron number on the oblate side with an even larger gap at the prolate deformation.
Conversely, the region with large $\delta S_{2n}$ around $N = 178$ is attributed to the 
large energy gap in the single-particle levels.

For protons, it is clear that the high $\delta S_{2p}$ at $Z = 120$ (with oblate ground-state) 
obtained with both parametrizations is due to the large energy gap in the single-particle levels at $Z = 120$.
On the other hand, a substantial energy gap at $Z = 124$ on the oblate side with the SLy5 parametrization is not seen in the the plot for $_{114}^{288}\mbox{Fl}_{174}$ nucleus.
Instead, the energy gap at $Z = 124$ appears at around $N = 180$
as seen in bottom single panel of Figure~\ref{fig: protons sp levels of Z = 124 N = 180}.

Apart from the above,
we also note that both Skyrme parametrizations show large energy gap in the single-particle energy levels at sphericity
for $Z = 126$ and $N = 184$. This reinforce the candidacy of $^{310}126_{184}$ as a doubly magic nucleus after $_{\;\;82}^{208}\mbox{Pb}_{126}$.

\subsection{Energy released in alpha-decay $Q_{\alpha}$ 
	\label{sec:results:alpha-decay Q-value}}

\begin{figure}[hbt!]\centering
	\includegraphics[width=1.0\textwidth]{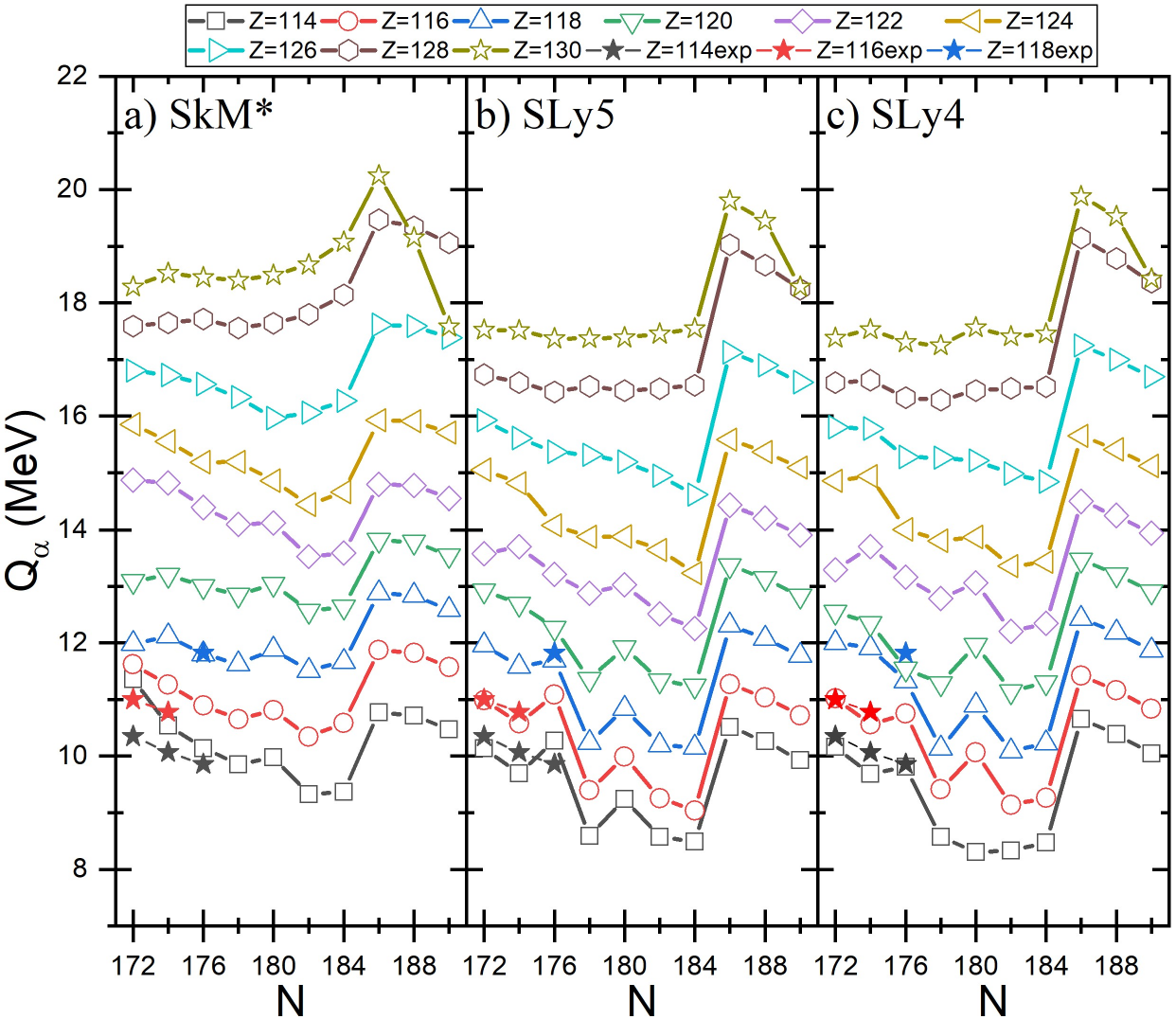}
	\caption{Alpha-decay energies, $Q_{\alpha}$-values (in MeV) for isotopic series $114 \leq Z\leq 126$ and $172 \leq N\leq 190$ for the SkM*, SLy5, and SLy4 Skyrme parametrizations as a function of $N$. Available experimental values taken from \cite{oganessian2015superheavy} and \cite{NNDC} are plotted in star shaped according to their designated colour.}
	\label{fig:Qalphavalues}
\end{figure}

The $Q_{\alpha}$ values calculated using equation (\ref{eq:Q-alpha}) with the ground-state energies as inputs, are tabulated in Table~\ref{tab:table3}
and plotted against neutron number in Figure~\ref{fig:Qalphavalues}.

In Figure~\ref{fig:Qalphavalues}, a peak can be seen at $N=$ 186 for all the isotopes 114 $\leq$ Z $\leq$ 130 in all  Skyrme parametrization plotted in panel (a), (b), and (c). 
This lends further support to $N=$ 184 (corresponding to the daughter nucleus after the alpha decay) as a strong candidate for spherical magic number, 
as was claimed by many previous studies (see Table~\ref{tab:list of magic candidates}).

Furthermore, it is interesting to point-out that a peak appears at $N= 180$  at lighter SH nuclei around $ 114 \le Z \le 122$ for SkM* and SLy5, and $ 116 \leq Z \leq 124$ for SLy4, except $Z=114$.
This suggest that $N=178$ is a candidate of deformed magic number, as was recently claimed by {\cite{wang2015systematic, cui2018alpha}}.
Figure~2 of Ref. \cite{cui2018alpha} showed magicity at $N=178$ and $N=184$ 
for isotopic series $Z=118, 119,$ and $120$.

On the other hand, the $Q_{\alpha}$ of even-even isotopes presented in Figure~4(b) in Ref.~\cite{wang2016correlations} 
showed possible magicity at $N=178$ for $116 \leq Z \leq 124$, and $N=184$ for $114 \leq Z \leq 126$. 
It was also suggested that $^{296}118$ and $^{298}120$ with both having $N = 178$
are much more stable in configuration with calculated oblate (deformed) shape. 

Comparing the calculated values to the limited experimental data available in the nuclear region,
it appears that the SkM* parametrization correctly reproduced the experimental downward trend for $172 \le N \le 176$ in the $Z = 114$ and $Z = 116$ isotope chains.

Nonetheless, the SLy5 and SLy4 predicted a peak in the alpha energy at $N = 176$ for selected isotopes, but this was not observed in the experiment.
The $N = 176$ $Q_{\alpha}$ peak obtained with the SLy5 and SLy4 parametrization 
is due to an energy gap in the single-particle spectra at $N = 174$ resulting in the emergence of a peak in the two-neutron separation energy differential. As for the SLy4 parametrization shows peaks at $N=174$ for $Z=124$ and $126$ indicating that $N=172$ to be magic.

The fact that the all three Skyrme parametrizations gave conflicting results with respect to the possibility of $N = 172$ and $174$ sub-magic numbers is interesting.
We note that {\cite{bender1999shell,bhuyan2012magic,malov2021landscape}} has also predicted a shell closure at these neutron numbers.
However, this point could only be verified through further experimental exploration.

\newpage
\subsection{Alpha-decay half-life $T_{1/2}$}
	\label{sec:results:alpha-decay half-life}

Using the $Q_{\alpha}$ values obtained with averaged pairing strengths (given in Table~\ref{tab:pairing strengths}), 
we calculated the logarithmic $\alpha$-decay half-lives ($\log_{10} T_{1/2}$) 
using various semi-empirical formulae as listed in Sec.~\ref{formulae}.
The calculated values using SkM*, SLy5 and SLy4  parametrizations are listed in Table~\ref{SkM* half-lives}, Table~\ref{SLy5 half-lives} and Table~\ref{SLy4 half-lives} , respectively.

We found that the nucleus with the longest $\log_{10} T_{1/2}$ within our samples to
be the $Z=114$ isotopes at $N=180$ (SLy4),  $N = 184$ (SLy5), and $N = 182$ (SkM*).

This finding reiterate the conclusion based on the single particle spectra (shell gaps) of Refs. \cite{rutz1997superheavy,bender1999shell} and
alpha decay half-life of Refs. \cite{berger2001superheavy,wang2015systematic}
which shows high stability for the $Z = 114, N = 184$ albeit a slight deviation in our case with the SLy4.

\begin{figure}[hbt!]
    \centering
	\includegraphics[width=1\textwidth]{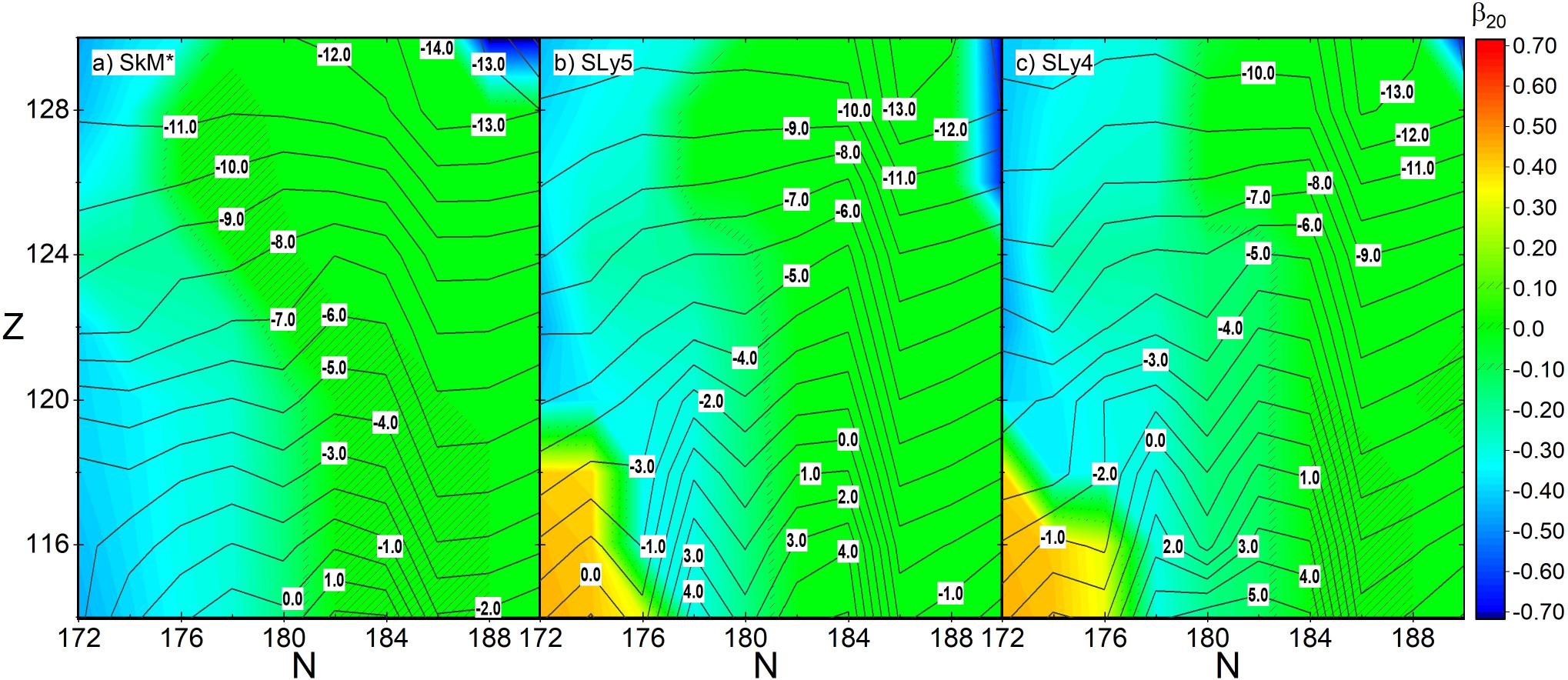}
	\caption{Shows overlapped between the ground-state $\beta_{20}$ given in color and the alpha-decay half-life $\log_{10} T_{1/2}$ from VS$1$ given in terms of contour lines. Nuclei along the isotone $N = 184$ exhibit spherical ground-state and therefore is a spherical magic number.}
	\label{fig: contour half-lives VS1}
\end{figure}

\begin{figure}[hbt!]\centering
	\includegraphics[width=1\textwidth]{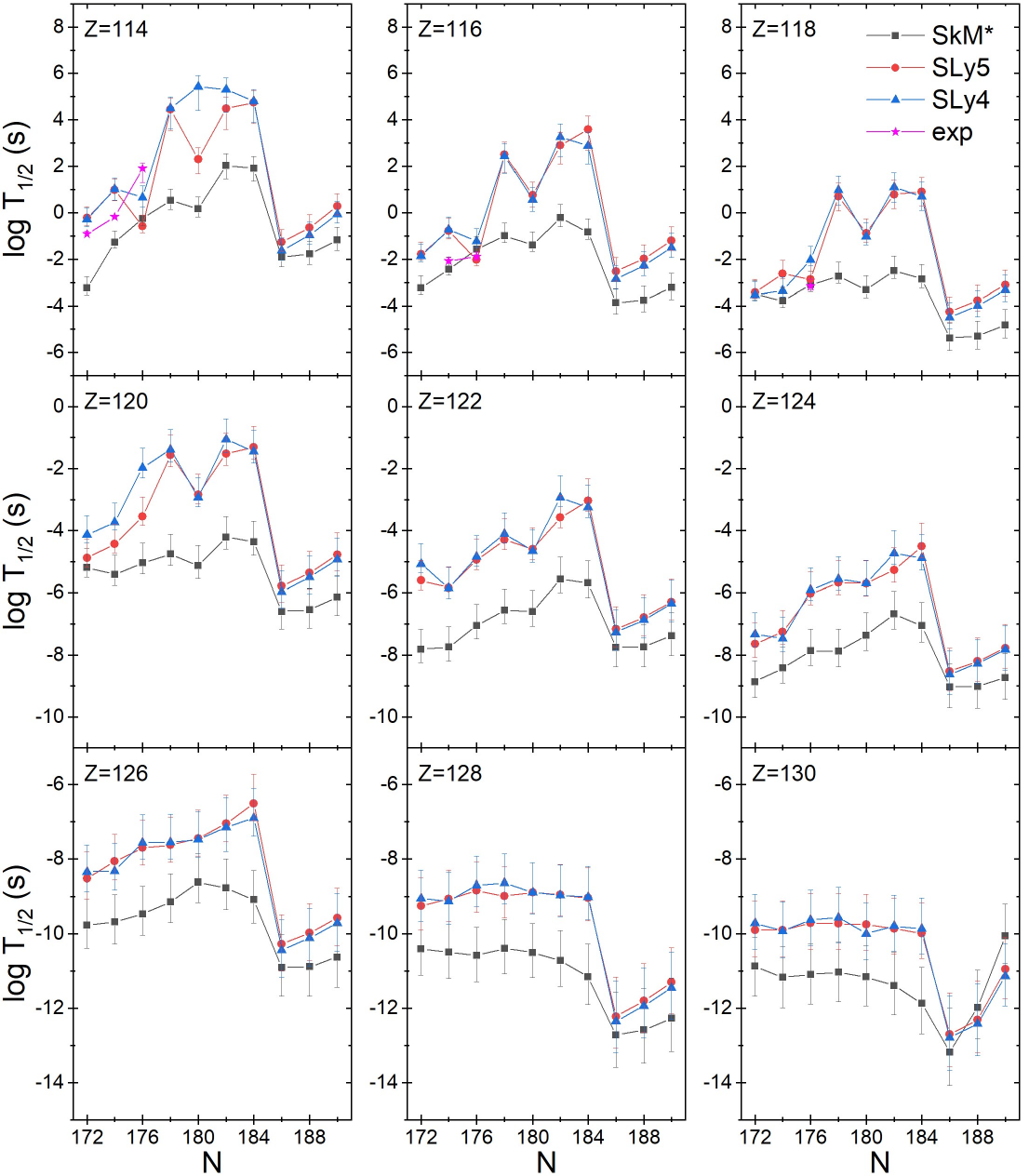}
	\caption{Shows the $\log_{10} T_{1/2}$ of all Skyrme parametrization are plotted from averaged of 8 different formulae against $(N)$ and the error bar are the maximum and minimum $T_{1/2}$ range for isotopic series 114 $\leq Z \leq$ 130 (172 $\leq N \leq$ 190). The experimental $\log T_{1/2}^{\alpha}$ are plotted using magenta star synmbol {\cite{NNDC}}.}
	\label{fig:halflives-neutron}
\end{figure}

Figure~\ref{fig: contour half-lives VS1} shows the evolution of our calculated $T_{1/2}$ based on the VS1 formula 
and superimposed with g.s.~$
\beta_{20}$ heatmap.
In panel (a) of the figure obtained with the SkM* parametrization, 
peaks in the $\log_{10} T_{1/2}$ are seen various neutron and proton numbers
namely at $N = 178$ with $Z = 114 - 122$, $N = 182$ with $Z = 114 - 124$ and $N = 184$ with $Z = 114 - 122$.
It is interesting to note here that the $N = 184$ loses its magic character at around $Z = 128$ and beyond.
This can be clearly seen in Figure~\ref{fig:halflives-neutron} which was plotted using the average $\log_{10} T_{1/2}$
of the eight semi-empirical formulas.
Contrast this to the results obtained with the SLy5 and SLy4 parametrization in panel (b) and (c) where $N = 182$ is not magic while 
$N = 184$ remains as a magic number
for all $Z$ numbers considered herein.
In addition to $N = 178$, a relatively high peak is seen at $N = 174$ with the SLy5 and SLy4 parametrizations.

Comparing the location of magic numbers with the g.s. deformation, 
we find that the $N = 178$ between $Z = 114 - 122$ 
forms a deformed (oblate) magic nuclei for all Skyrme parametrizations.
For $N = 182$, we found that the $Z = 114 - 124$ forms a (spherical) shell closure with the SkM* parametrizations, 
whereas $N = 184$ is spherical for all Skyrme parametrizations. 
In panel (b) and (c) obtained with the SLy5 and SLy4 parametrization, 
we found that $N=174$ could be a candidate for a deformed (prolate) magic number.

In what follows, we keep our analyses to quantities which are taken
as average of half lives obtained with the 8 semi-empirical formulas discussed in Sec~\ref{formulae}.
Further discussion on the differences between results of the different formulas
are provided in the \ref{sec:appendix 1}.

These averaged $\log_{10} T_{1/2}$ values are plotted as a function of $N$ in Figure~\ref{fig:halflives-neutron}.
The largest negative and positive differences with respect to the averaged values are plotted as error bars to indicate the uncertainties in semi-empirical formulae.

The $\log_{10} T_{1/2}$ systematic plots in Figure~\ref{fig:halflives-neutron} compliments the observation from Figure~\ref{fig: contour half-lives VS1} by showing magic candidate peaks at $N= 178$ and $182$ or $184$, 
where there appears a sharp decrease in the alpha decay $T_{1/2}$ when going from $N = 184$ to $N = 186$ 
for all Skyrme parametrizations. 
The experimental $\log_{10} T_{1/2}$ for $Z= 114 - 118$ are shown in the same figure. 
Interestingly, SkM* parametrization $\log_{10}T_{1/2}$ trends are consistent with $\log_{10} T_{1/2}^{exp.}$, 
which suggest that $N=174$ $(Z=114-118)$ is not a magic number candidate 
contrary to the findings obtained with the SLy5 and SLy4 parametrizations 
and also in the recent study of by Malov et al. \cite{malov2021landscape}.

It is also interesting to note that our calculations showed a rather different
pattern in alpha-decay half-life for $N \ge 188$ as compared to Figure~4 of {\cite{ghodsi2020alpha}}.
In {\cite{ghodsi2020alpha}}, the authors predicted
that the alpha-decay half-life of $Z = 126$ with $N = 188$ nucleus to be higher
than the half-life of the doubly magic nucleus $^{310}126_{184}$ by about 2 orders of magnitude.
In our calculations, we found that alpha-decay half-life for $186 \le N \le 190$ is always lower than its neighbour $N = 184$ for any $Z$ isotopes, except for $Z = 130$ obtained using SkM* parametrization.

\begin{figure}[hbt!]\centering
	\includegraphics[width=\textwidth]{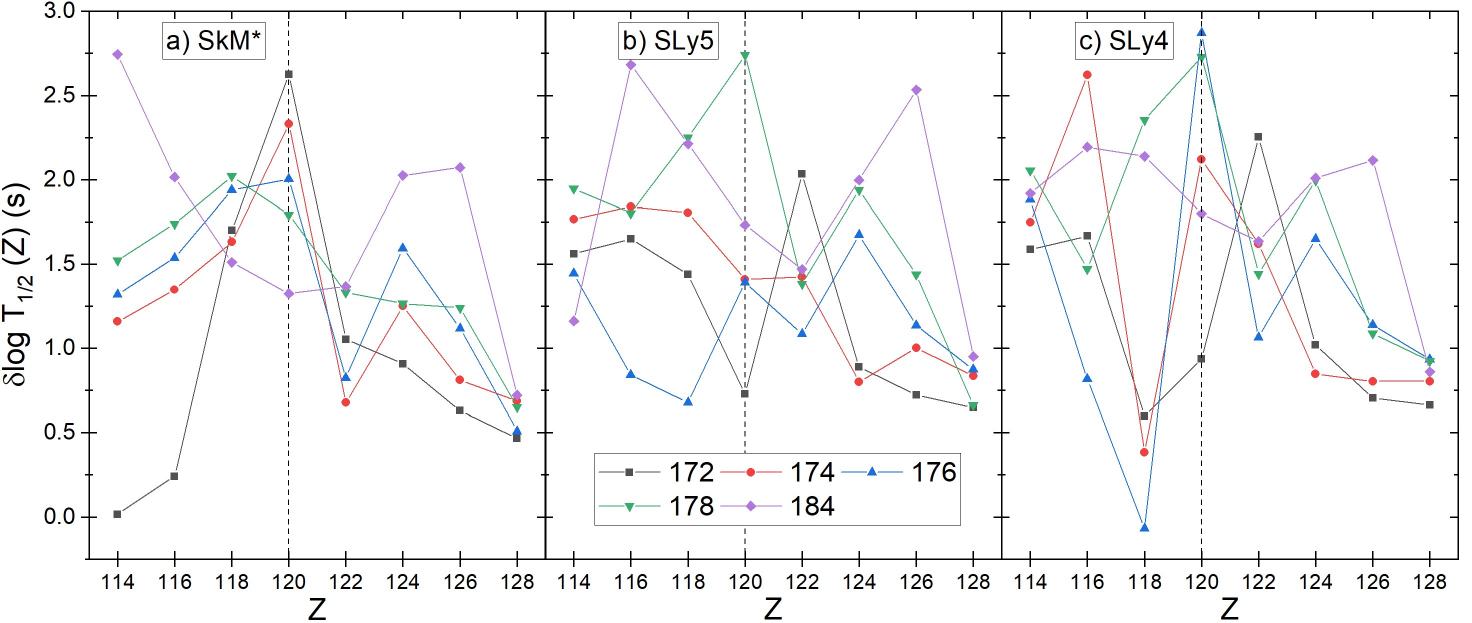}
	\caption{Shows difference in alpha-decay half-life $\log_{10}T_{1/2}$ as a function of $N$ for proton number $Z = 120$ and function of $Z$ for neutron number $N = 178$ obtained with all Skyrme parametrization.}
	\label{fig: delta half-life vs neutron}
\end{figure}

We now move on to discuss the magic number candidates for protons. To this end, we plotted the difference in the half-life of $\log_{10} T_{1/2}$, $(\delta \: \{\mbox{log}_{10} T_{1/2} (Z)\})$, 
as a function of proton number $Z$ as shown in Figure~\ref{fig: delta half-life vs neutron} 
with fixed $N = 172 - 178$, and $184$ using the equation:
\begin{equation}
	\delta \: \{\mbox{log}_{10} T_{1/2} (Z)\} = \mbox{log}_{10} \: T_{1/2} (Z) - \mbox{log}_{10} \: T_{1/2} (Z + 2).
	\label{eq: delta half-life}
\end{equation}

In panel (a) of Figure~\ref{fig: delta half-life vs neutron} obtained with the SkM* parameterization, a sharp peak appears in $\delta \: \{\mbox{log}_{10} T_{1/2} (Z)\}$ at $Z = 120$ for $N = 172$ and $174$, and another small peak at $Z = 124$ for $N = 174$ and $176$.
This is associated to the region with high and moderate $\delta S_{2p}$ in panel (c) of Figure~\ref{fig:contour s2n s2p} at the corresponding $Z$ and $N$ numbers. The peak appears instead at $Z = 126$ and with possibly another higher peak around $Z = 114$ corresponding to a small region of high $\mbox{log}~T_{1/2}$ in Figure~\ref{fig: contour half-lives VS1}. 

Next, in the similar plot for the SLy5 and SLy4 parametrization plotted in panel (b) and (c)
the peaks in $(\delta \: \{\mbox{log}_{10} T_{1/2} (Z)\})$ appears at $Z = 120$ similar to what was obtained with the SkM* parametrization but at different neutron number namely $N = 176$ and $178$ for SLy5, and $ 174 \leq Z \leq 178 $ for SLy4. 
For $N = 184$, no peak is observed at $Z = 120$.
Instead, a slight peak appears at $Z = 116$ and a more pronounced peak appears at $Z = 126$.

\section{Conclusion}\label{conclusion}

In this article, we presented our calculations based on the Skyrme Hartree-Fock-plus-BCS framework for superheavy nuclei in the region of $112 \le Z \le 130$ and $172 \le N \le 190$. The aim is to identify the candidates for neutron and proton magic numbers by simultaneously evaluating the single-particle properties via two-nucleon separation energy differential and single-particle energy levels, and the bulk properties via $Q_{\alpha}$ energy and alpha-decay half-life $\mbox{log}_{10} T_{1/2}$.

We found that $N = 178$ to be a candidate for neutron deformed magic number around $114 \le Z \le 118$, but not for $N=174$.
This finding is supported by the emergence of peaks in the $\mbox{log}_{10} T_{1/2}$ as a function of $N$,
and the two-neutron separation energy differential $\delta S_{2n}$.
Emergence of this deformed magic number is supported by a significant energy gap in the single-particle energy levels with an oblate ground-state shape. 
While results obtained with the SLy5 and SLy4 show a potential deformed magic number at N = 174, the actual magic character at N = 174 may be unlikely based on the trend in the experimental alpha decay energies and half-life.

For protons, we found that $Z = 120$ and $124$ to be a candidate for deformed magic number at
around $N = 172 - 176$ (with the SkM*), $N= 176 - 178$ (with the SLy5), and $N= 174 - 178$ (with the SLy4).
This was identified from the analyses of two-proton separation energy differential $\delta S_{2p}$ and reconfirmed from the 
analyses of the variation in the alpha-decay half-life $\delta \: \mbox{log} T_{1/2}$ as a function of $Z$ number.

In addition to deformed magic numbers, we have also found simultaneous peaks 
in the $\delta S_{2n}$ and $\delta S_{2p}$ at $N = 184$ and $Z = 126$, respectively.
Coupled with the fact that the ground-state deformation was found at sphericity,
we arrived at the same conclusion as was found before, that $^{310}126_{184}$
is a doubly magic nucleus.

\begin{figure}[hbt!]\centering
	\includegraphics[width=0.70\textwidth]{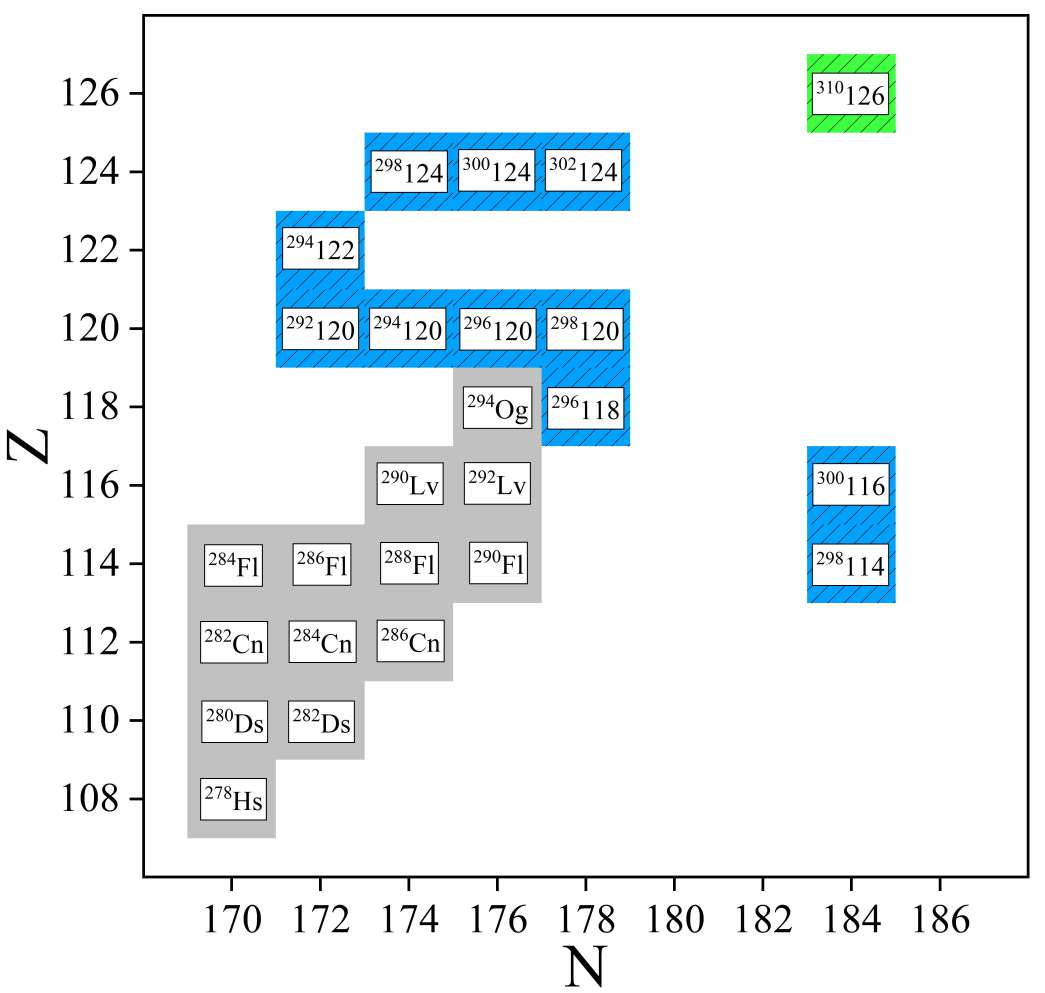}
	\caption{Shows even-even nuclei in the superheavy region of nuclear chart. 
		The boxes in grey represents known synthesized nuclei while
		boxes in blue are suggested candidates nuclei for exploring the superheavy region towards 
		the doubly magic $^{310}126$ in green.}
	\label{fig: nuclear chart synthesis}
\end{figure}

The information presented herein is of interest in the
effort to synthesize new superheavy elements. 
Figure-\ref{fig: nuclear chart synthesis} illustrates the locations for nuclei that could be of interest within the superheavy region.
In particular are nuclei around (i) $Z = 114 - 116$ with $N = 184$, (ii) $Z = 118$ with $N = 178$ 
(iii) $Z = 120$ around $N = 172 - 178$, (iv) $Z = 122 - 124$ with $N = 172 - 178$.

\section*{Acknowledgements}

The authors would like to acknowledge the Ministry of Education Malaysia for the financial support through the Fundamental Research Grant Scheme (FRGS/1/2018/ST/G02/UTM/02/6), and UTM (R.J130000.7854.5F028). \\

 \newpage
\begingroup
\begin{longtable}{>{\centering\arraybackslash}p{1.0cm}>{\centering\arraybackslash}p{1.0cm}>{\centering\arraybackslash}p{1.0cm}>{\centering\arraybackslash}p{1.0cm}>{\centering\arraybackslash}p{1.0cm}>{\centering\arraybackslash}p{1.0cm}>{\centering\arraybackslash}p{1.0cm}>{\centering\arraybackslash}p{1.0cm}>{\centering\arraybackslash}p{1.0cm}}
	\caption{\label{tab:binding energy per nucleon}
            Binding energy per nucleon $BE/A$ obtained in current work with the SkM*, SLy5 and SLy4 parametrizations
            and compared with the results obtained from Ref.~\cite{karim2019self} using the Skyrme HFB framework.} \\ \toprule
	\multirow{2}{*}{$Z$} & \multirow{2}{*}{$N$} & \multirow{2}{*}{$A$} &           
       \multicolumn{6}{c}{BE/A}  \\
        \cmidrule{4-9}
        & & & {SkM*} & {SLy5} & {SLy4} & {SkM* (HFB)} & {SLy5 (HFB)} & {SLy4 (HFB)}  \\ 
	\colrule 
	\endfirsthead
	\multicolumn{7}{r}{\tablename\ \thetable\ -- \textit{Continued from previous page}} \\
	\toprule 
	\multirow{2}{*}{$Z$} & \multirow{2}{*}{$N$} & \multirow{2}{*}{$A$} &           
       \multicolumn{6}{c}{BE/A}  \\
        \cmidrule{4-9}
        & & & {SkM*} & {SLy5} & {SLy4} & {SkM* (HFB)} & {SLy5 (HFB)} & {SLy4 (HFB)}  \\  
	\colrule 
	\endhead 
	\hline \multicolumn{9}{r}{\tablename\ \thetable\ -- \textit{Continued on next page}} \\
	\endfoot
	\hline 
	\endlastfoot
112 & 170 & 282 & 7.147 & 7.151 & 7.148 & 7.133    & 7.123    & 7.134    \\
112 & 172 & 284 & 7.142 & 7.145 & 7.142 & 7.128    & 7.119    & 7.128    \\
112 & 174 & 286 & 7.139 & 7.138 & 7.134 & --       & --       & --       \\
112 & 176 & 288 & 7.134 & 7.126 & 7.124 & 7.117    & 7.098    & 7.108    \\
112 & 178 & 290 & 7.128 & 7.119 & 7.112 & --       & --       & --       \\
112 & 180 & 292 & 7.120 & 7.106 & 7.103 & --       & --       & --       \\
112 & 182 & 294 & 7.111 & 7.095 & 7.091 & --       & --       & --       \\
112 & 184 & 296 & 7.101 & 7.082 & 7.077 & --       & --       & --       \\
\end{longtable}
\endgroup

\begingroup
\begin{longtable}{>{\centering\arraybackslash}p{1.0cm}>{\centering\arraybackslash}p{1.0cm}>{\centering\arraybackslash}p{1.0cm}>{\centering\arraybackslash}p{1.0cm}>{\centering\arraybackslash}p{1.0cm}>{\centering\arraybackslash}p{1.0cm}>{\centering\arraybackslash}p{1.0cm}>{\centering\arraybackslash}p{1.0cm}>{\centering\arraybackslash}p{1.0cm}}
	\caption{Calculated root-mean-square radius $r_{rms}$ for Skyrme parameters of SkM*, SLy5, and SLy4. \label{tab: root-mean-square radii}} \\ \toprule
	\multirow{2}{*}{$Z$} & \multirow{2}{*}{$N$} & \multirow{2}{*}{$A$} &           
       \multicolumn{6}{c}{$r_{rms}$}  \\
        \cmidrule{4-9}
        & & & {SkM*} & {SLy5} & {SLy4} & {SkM* (HFB)} & {SLy5 (HFB)} & {SLy4 (HFB)}   \\ 
	\colrule 
	\endfirsthead
	\multicolumn{9}{r}{\tablename\ \thetable\ -- \textit{Continued from previous page}} \\
	\toprule 
	\multirow{2}{*}{$Z$} & \multirow{2}{*}{$N$} & \multirow{2}{*}{$A$} &           
       \multicolumn{6}{c}{$r_{rms}$}  \\
        \cmidrule{4-9}
        & & & {SkM*} & {SLy5} & {SLy4} & {SkM* (HFB)} & {SLy5 (HFB)} & {SLy4 (HFB)}  \\  
	\colrule 
	\endhead 
	\hline \multicolumn{9}{r}{\tablename\ \thetable\ -- \textit{Continued on next page}} \\
	\endfoot
	\hline 
	\endlastfoot
112 & 170 & 282 & 6.223  & 6.206 & 6.214 & --       & --       & --        \\
112 & 172 & 284 & 6.214  & 6.213 & 6.220 & --       & --       & --       \\
112 & 174 & 286 & 6.224  & 6.224 & 6.231 & --       & --       & --       \\
112 & 176 & 288 & 6.233  & 6.228 & 6.234 & --       & --       & --       \\
112 & 178 & 290 & 6.245  & 6.236 & 6.241 & --       & --       & --       \\
112 & 180 & 292 & 6.254  & 6.245 & 6.251 & --       & --       & --       \\
112 & 182 & 294 & 6.265  & 6.256 & 6.262 & --       & --       & --       \\
112 & 184 & 296 & 6.278  & 6.267 & 6.273 & --       & --       & --       \\
\end{longtable}
\endgroup

\begingroup
\begin{longtable}{>{\centering\arraybackslash}p{1.0cm}>{\centering\arraybackslash}p{1.0cm}>{\centering\arraybackslash}p{1.0cm}>{\centering\arraybackslash}p{1.0cm}>{\centering\arraybackslash}p{1.0cm}>{\centering\arraybackslash}p{1.0cm}>{\centering\arraybackslash}p{1.0cm}>{\centering\arraybackslash}p{1.0cm}>{\centering\arraybackslash}p{1.0cm}}
	\caption{Calculated charge radius $r_{p}$ for Skyrme parameters of SkM*, SLy5, and SLy4 \label{tab: charge radius}} \\ \toprule
	\multirow{2}{*}{$Z$} & \multirow{2}{*}{$N$} & \multirow{2}{*}{$A$} &           
       \multicolumn{6}{c}{$r_{p}$}  \\
        \cmidrule{4-9}
        & & & {SkM*} & {SLy5} & {SLy4} & {SkM* (HFB)} & {SLy5 (HFB)} & {SLy4 (HFB)}   \\  
	\colrule 
	\endfirsthead
	\multicolumn{9}{r}{\tablename\ \thetable\ -- \textit{Continued from previous page}} \\
	\toprule 
	\multirow{2}{*}{$Z$} & \multirow{2}{*}{$N$} & \multirow{2}{*}{$A$} &           
       \multicolumn{6}{c}{$r_{p}$}  \\
        \cmidrule{4-9}
       & & & {SkM*} & {SLy5} & {SLy4} & {SkM* (HFB)} & {SLy5 (HFB)} & {SLy4 (HFB)}   \\ 
	\colrule 
	\endhead 
	\hline \multicolumn{9}{r}{\tablename\ \thetable\ -- \textit{Continued on next page}} \\
	\endfoot
	\hline 
	\endlastfoot
112 & 170 & 282 & 6.151    & 6.135 & 6.145 & --       & --       & --    \\
112 & 172 & 284 & 6.138    & 6.138 & 6.147  & --       & --       & --      \\
112 & 174 & 286 & 6.142    & 6.145 & 6.153  & --       & --       & --       \\
112 & 176 & 288 & 6.146    & 6.144 & 6.152 & --  & --       & --   \\
112 & 178 & 290 & 6.152    & 6.146 & 6.153 & --   & --       & --  \\
112 & 180 & 292 & 6.156    & 6.149 & 6.157 & --   & --       & --  \\
112 & 182 & 294 & 6.160    & 6.154 & 6.162 & --   & --       & --  \\
112 & 184 & 296 & 6.166    & 6.159 & 6.167 & --   & --       & --  \\
\end{longtable}
\endgroup

\begingroup
\begin{longtable}{>{\centering\arraybackslash}p{1cm}>{\centering\arraybackslash}p{1cm}>{\centering\arraybackslash}p{1cm}>{\centering\arraybackslash}p{1.8cm}>{\centering\arraybackslash}p{1.8cm}>{\centering\arraybackslash}p{1.8cm}>{\centering\arraybackslash}p{1.8cm}}
	\caption{\label{tab:table3}Calculated $Q_\alpha$-values (MeV) with Skyrme parametrization SkM* ($Q_{SkM^*}$), SLy5 ($Q_{SLy5}$) and SLy4 ($Q_{SLy4}$) with the averaged of all Madland's pairing strength, and experimental $Q_\alpha$-value ($Q_\alpha^{exp.}$) are taken from {\cite{oganessian2015superheavy} and \cite{NNDC}}.} \\ \toprule
	Z   & N   &  A & $Q_{SkM*}^{Ave}$ &  $Q_{SLy5}^{Ave}$ & $Q_{SLy4}^{Ave}$ & $Q_\alpha^{exp.}$ \\  \colrule
	\endfirsthead
	\multicolumn{7}{r}{\tablename\ \thetable\ -- \textit{Continued from previous page}} \\
	\toprule 
	Z   & N   & A   & $Q_{SkM*}^{Ave}$  & $Q_{SLy5}^{Ave}$ & $Q_{SLy4}^{Ave}$ & $Q_\alpha^{exp.}$ \\ 
	\colrule 
	\endhead 
	\botrule \multicolumn{7}{r}{\tablename\ \thetable\ -- \textit{Continued on next page}} \\
	\endfoot
	\hline 
	\endlastfoot
	114 & 172 & 286 & 11.36      & 10.14       & 10.16       & 10.35  \\
114 & 174 & 288 & 10.54      & 9.70        & 9.68        & 10.07  \\
114 & 176 & 290 & 10.14      & 10.27       & 9.81        & 9.86   \\
114 & 178 & 292 & 9.85       & 8.59        & 8.57        & --     \\
114 & 180 & 294 & 9.98       & 9.24        & 8.31        & --     \\
114 & 182 & 296 & 9.33       & 8.57        & 8.34        & --     \\
114 & 184 & 298 & 9.37       & 8.49        & 8.48        & --     \\
114 & 186 & 300 & 10.77      & 10.51       & 10.66       & --     \\
\end{longtable}
\endgroup

\begingroup
\begin{longtable}{>{\centering\arraybackslash}p{0.4cm}>{\centering\arraybackslash}p{0.4cm}>{\centering\arraybackslash}p{0.4cm}>{\centering\arraybackslash}p{1.0cm}>{\centering\arraybackslash}p{1.0cm}>{\centering\arraybackslash}p{1.0cm}>{\centering\arraybackslash}p{1.0cm}>{\centering\arraybackslash}p{1.0cm}>{\centering\arraybackslash}p{1.0cm}>{\centering\arraybackslash}p{1.0cm}>{\centering\arraybackslash}p{1.0cm}>{\centering\arraybackslash}p{1.0cm}}
	\caption{\label{SkM* half-lives} Comparison of calculated logarithmic $\alpha$-decay half-lives with Skyrme SkM* Skyrme parametrization $\log T_{1/2}^{SkM^*}$ (s) together with experimental half-lives $\log T_{1/2}^{exp.}$ (s) taken from {\cite{oganessian2015superheavy} and \cite{NNDC}}.
		\label{tab:half-life table SkM}} \\ \toprule
	Z   & N   & A   & $T_{1/2}^{VSS1}$    & $T_{1/2}^{VSS2}$    & $T_{1/2}^{PS}$      & $T_{1/2}^{R}$    & $T_{1/2}^{mR}$    & $T_{1/2}^{BF}$      & $T_{1/2}^{MB1}$     & $T_{1/2}^{MB2}$     & $T_{1/2}^{exp.}$    \\  \midrule
	\endfirsthead
	\multicolumn{12}{r}{\tablename\ \thetable\ -- \textit{\Tstrut Continued from previous page}} \\
	\toprule 
	Z   & N   & A   & $T_{1/2}^{VSS1}$    & $T_{1/2}^{VSS2}$    & $T_{1/2}^{PS}$      & $T_{1/2}^{R}$    & $T_{1/2}^{mR}$    & $T_{1/2}^{BF}$      & $T_{1/2}^{MB1}$     & $T_{1/2}^{MB2}$     & $T_{1/2}^{exp.}$    \\  \midrule
	\endhead 
	\midrule \multicolumn{12}{r}{\tablename\ \thetable\ -- \textit{\Tstrut Continued on next page}} \\
	\endfoot 
	\bottomrule
	\endlastfoot 
114 & 172 & 286 & -3.29  & -2.74  & -2.90  & -3.53  & -3.48  & -3.36  & -3.17  & -3.37  & -0.92 \\
114 & 174 & 288 & -1.24  & -0.79  & -0.92  & -1.51  & -1.47  & -1.52  & -1.46  & -1.32  & -0.18 \\
114 & 176 & 290 & -0.15  & 0.25   & 0.14   & -0.46  & -0.42  & -0.55  & -0.55  & -0.24  & 1.90  \\
114 & 178 & 292 & 0.67   & 1.03   & 0.93   & 0.32   & 0.36   & 0.19   & 0.13   & 0.59   & --     \\
114 & 180 & 294 & 0.30   & 0.68   & 0.58   & -0.08  & -0.04  & -0.14  & -0.17  & 0.22   & --     \\
114 & 182 & 296 & 2.27   & 2.55   & 2.48   & 1.85   & 1.88   & 1.62   & 1.46   & 2.18   & --     \\
114 & 184 & 298 & 2.14   & 2.43   & 2.36   & 1.69   & 1.72   & 1.50   & 1.36   & 2.06   & --    \\
114 & 186 & 300 & -1.85  & -1.37  & -1.50  & -2.34  & -2.29  & -2.06  & -1.96  & -1.93  & --     \\
\end{longtable}
\endgroup
\normalsize

\begingroup
\begin{longtable}{>{\centering\arraybackslash}p{0.4cm}>{\centering\arraybackslash}p{0.4cm}>{\centering\arraybackslash}p{0.4cm}>{\centering\arraybackslash}p{1.0cm}>{\centering\arraybackslash}p{1.0cm}>{\centering\arraybackslash}p{1.0cm}>{\centering\arraybackslash}p{1.0cm}>{\centering\arraybackslash}p{1.0cm}>{\centering\arraybackslash}p{1.0cm}>{\centering\arraybackslash}p{1.0cm}>{\centering\arraybackslash}p{1.0cm}>{\centering\arraybackslash}p{1.0cm}}
	\caption{\label{SLy5 half-lives}Comparison of calculated logarithmic $\alpha$-decay half-lives with Skyrme SLy5 parametrization $\log T_{1/2}^{SLy5}$ (s) together with experimental half-lives $ \log T_{1/2}^{exp.}$ (s) taken from {\cite{oganessian2015superheavy} and \cite{NNDC}}.
		\label{tab:half-life table SLy5}}
	\\ \toprule
	Z   & N   & A   & $T_{1/2}^{VSS1}$    & $T_{1/2}^{VSS2}$    & $T_{1/2}^{PS}$      & $T_{1/2}^{R}$    & $T_{1/2}^{mR}$    & $T_{1/2}^{BF}$      & $T_{1/2}^{MB1}$     & $T_{1/2}^{MB2}$     & $T_{1/2}^{exp.}$    \\  \colrule
	\endfirsthead
	\multicolumn{12}{r}{\tablename\ \thetable\ -- \textit{\Tstrut Continued from previous page}} \\
	\toprule 
	Z   & N   & A   & $T_{1/2}^{VSS1}$    & $T_{1/2}^{VSS2}$    & $T_{1/2}^{PS}$      & $T_{1/2}^{R}$    & $T_{1/2}^{mR}$    & $T_{1/2}^{BF}$      & $T_{1/2}^{MB1}$     & $T_{1/2}^{MB2}$     & $T_{1/2}^{exp.}$    \\  \colrule
	\endhead 
	\colrule \multicolumn{12}{r}{\tablename\ \thetable\ -- \textit{\Tstrut Continued on next page}} \\
	\endfoot 
	\bottomrule
	\endlastfoot 
	114 & 172 & 286 & -0.15  & 0.25   & 0.14   & -0.38  & -0.34  & -0.54  & -0.55  & -0.23  & -0.92 \\
	114 & 174 & 288 & 1.13   & 1.47   & 1.38   & 0.85   & 0.89   & 0.60   & 0.51   & 1.04   & -0.18 \\
	114 & 176 & 290 & -0.51  & -0.09  & -0.21  & -0.82  & -0.78  & -0.87  & -0.85  & -0.59  & 1.90  \\
	114 & 178 & 292 & 4.77   & 4.93   & 4.90   & 4.42   & 4.45   & 3.85   & 3.54   & 4.68   & --     \\
	114 & 180 & 294 & 2.54   & 2.81   & 2.75   & 2.16   & 2.20   & 1.86   & 1.69   & 2.46   & --     \\
	114 & 182 & 296 & 4.82   & 4.98   & 4.96   & 4.40   & 4.44   & 3.90   & 3.59   & 4.73   & --     \\
	114 & 184 & 298 & 5.12   & 5.26   & 5.24   & 4.66   & 4.69   & 4.16   & 3.84   & 5.03   & --     \\
	114 & 186 & 300 & -1.17  & -0.72  & -0.84  & -1.66  & -1.61  & -1.46  & -1.40  & -1.25  & --     \\
\end{longtable}
\endgroup

\begingroup
\begin{longtable}{>{\centering\arraybackslash}p{0.4cm}>{\centering\arraybackslash}p{0.4cm}>{\centering\arraybackslash}p{0.4cm}>{\centering\arraybackslash}p{1.0cm}>{\centering\arraybackslash}p{1.0cm}>{\centering\arraybackslash}p{1.0cm}>{\centering\arraybackslash}p{1.0cm}>{\centering\arraybackslash}p{1.0cm}>{\centering\arraybackslash}p{1.0cm}>{\centering\arraybackslash}p{1.0cm}>{\centering\arraybackslash}p{1.0cm}>{\centering\arraybackslash}p{1.0cm}}
	\caption{\label{SLy4 half-lives} Comparison of calculated logarithmic $\alpha$-decay half-lives with Skyrme SLy4 Skyrme parametrization $\log T_{1/2}^{SLy4}$ (s) together with experimental half-lives $\log T_{1/2}^{exp.}$ (s) taken from {\cite{oganessian2015superheavy} and \cite{NNDC}}.
		\label{tab:half-life table SLy4}} \\ \toprule
	Z   & N   & A   & $T_{1/2}^{VSS1}$    & $T_{1/2}^{VSS2}$    & $T_{1/2}^{PS}$      & $T_{1/2}^{R}$    & $T_{1/2}^{mR}$    & $T_{1/2}^{BF}$      & $T_{1/2}^{MB1}$     & $T_{1/2}^{MB2}$     & $T_{1/2}^{exp.}$    \\  \midrule
	\endfirsthead
	\multicolumn{12}{r}{\tablename\ \thetable\ -- \textit{\Tstrut Continued from previous page}} \\
	\toprule 
	Z   & N   & A   & $T_{1/2}^{VSS1}$    & $T_{1/2}^{VSS2}$    & $T_{1/2}^{PS}$      & $T_{1/2}^{R}$    & $T_{1/2}^{mR}$    & $T_{1/2}^{BF}$      & $T_{1/2}^{MB1}$     & $T_{1/2}^{MB2}$     & $T_{1/2}^{exp.}$    \\  \midrule
	\endhead 
	\midrule \multicolumn{12}{r}{\tablename\ \thetable\ -- \textit{\Tstrut Continued on next page}} \\
	\endfoot 
	\bottomrule
	\endlastfoot 
114 & 172 & 286 & -0.21  & 0.19   & 0.08   & -0.44  & -0.40  & -0.60  & -0.60  & -0.29  & -0.92  \\
114 & 174 & 288 & 1.17   & 1.51   & 1.42   & 0.90   & 0.93   & 0.63   & 0.55   & 1.08   & -0.18  \\
114 & 176 & 290 & 0.80   & 1.16   & 1.06   & 0.49   & 0.53   & 0.30   & 0.24   & 0.72   & 1.90  \\
114 & 178 & 292 & 4.83   & 4.99   & 4.96   & 4.48   & 4.52   & 3.91   & 3.60   & 4.74   & --  \\
114 & 180 & 294 & 5.80   & 5.91   & 5.90   & 5.41   & 5.45   & 4.78   & 4.40   & 5.71   & --  \\
114 & 182 & 296 & 5.70   & 5.82   & 5.81   & 5.28   & 5.31   & 4.69   & 4.32   & 5.61   & --  \\
114 & 184 & 298 & 5.17   & 5.31   & 5.29   & 4.71   & 4.75   & 4.21   & 3.88   & 5.08   & --  \\
114 & 186 & 300 & -1.55  & -1.08  & -1.22  & -2.04  & -2.00  & -1.80  & -1.72  & -1.63  & --  \\
\end{longtable}
\endgroup
\normalsize

\appendix
\section{Comparison of alpha-decay half-lives from various equations}
\label{sec:appendix 1}
\begin{figure}[hbt!]\centering
	\includegraphics[width=1 \textwidth]{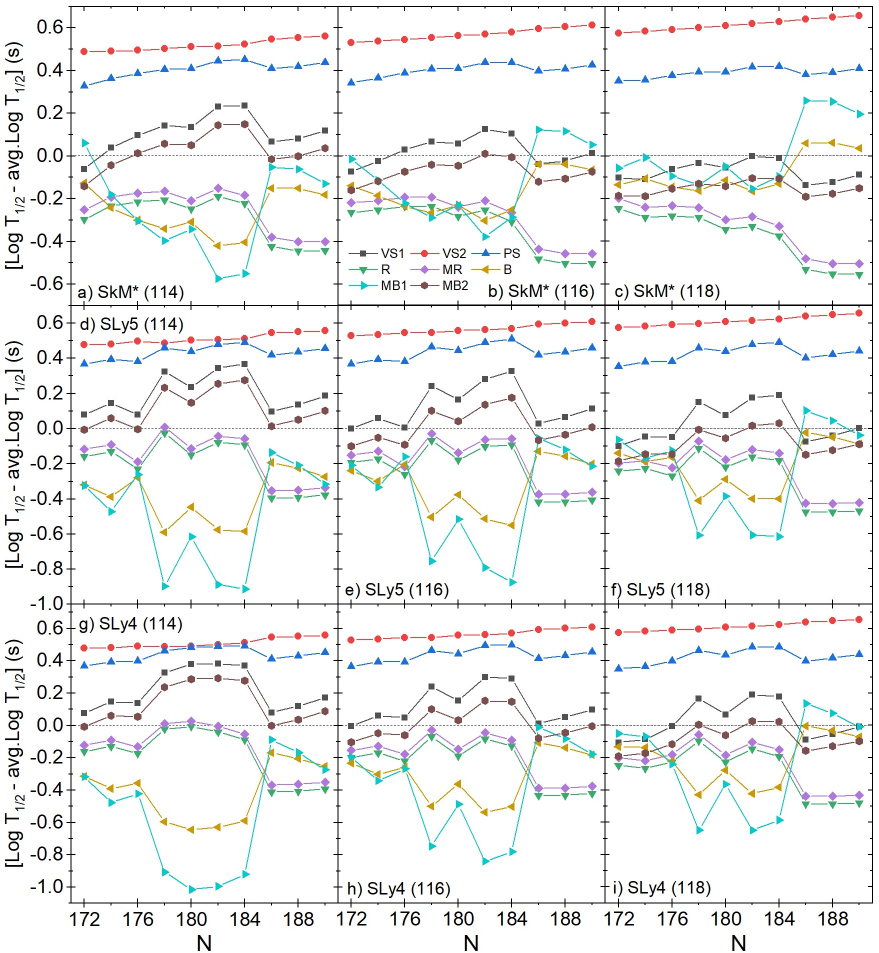}
	\caption{Analysis of 8 different formulae for isotopes $Z=114 - 118$ for all Skyrme paramterization (SkM*, SLy5, and SLy5).}
	\label{fig: 114-118 analysis}
\end{figure}

\begin{figure}[hbt!]\centering
	\includegraphics[width=1 \textwidth]{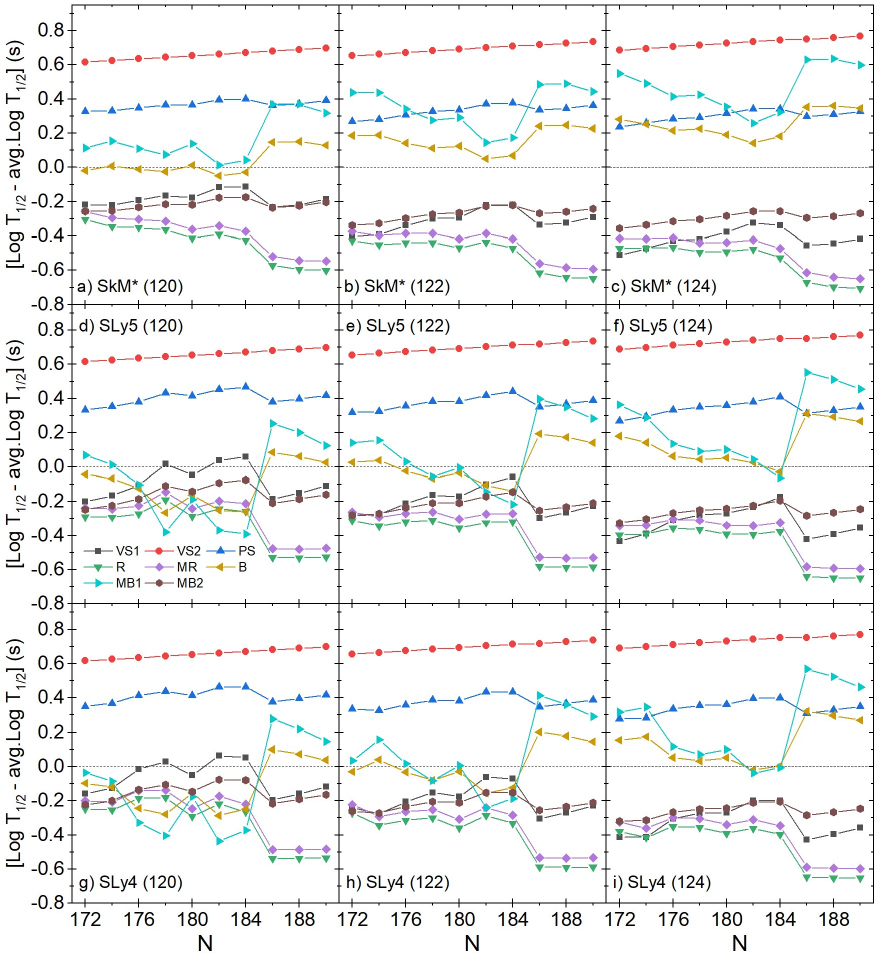}
	\caption{Analysis of 8 different formulae for isotopes $Z=120 - 124$ for all Skyrme paramterization (SkM*, SLy5, and SLy5).}
	\label{fig: 120-124 analysis}
\end{figure}

\begin{figure}[hbt!]\centering
	\includegraphics[width=1 \textwidth]{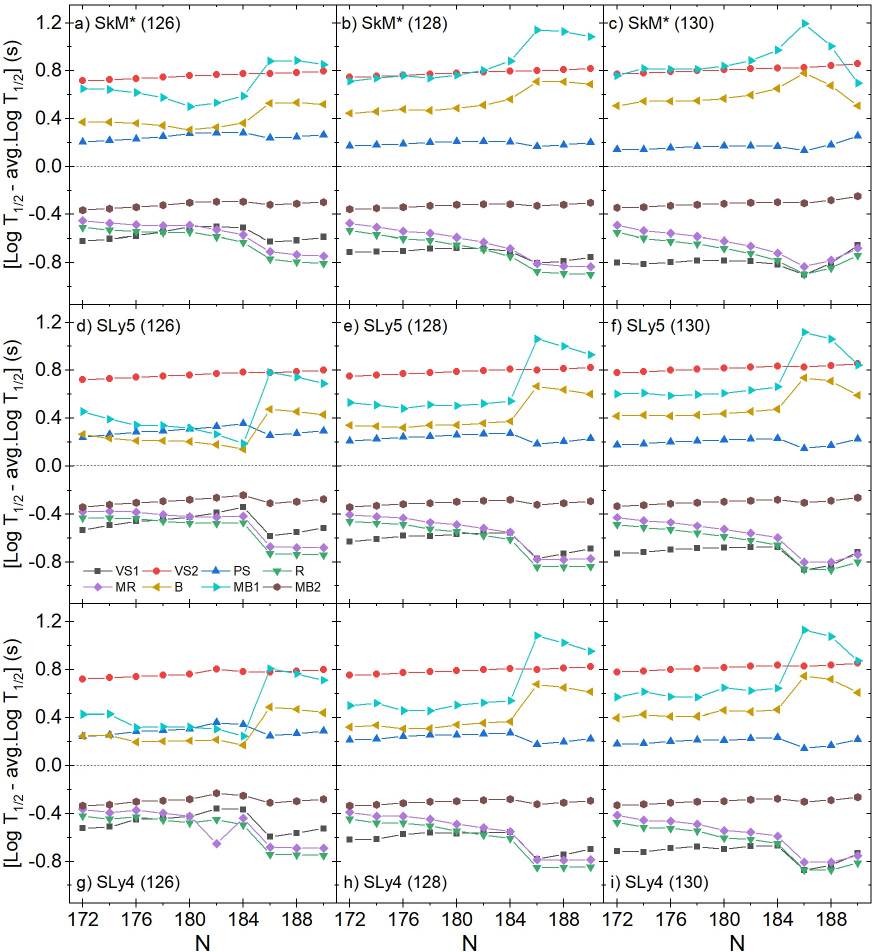}
	\caption{Analysis of 8 different formulae for isotopes $Z=126 - 130$ for all Skyrme paramterization (SkM*, SLy5, and SLy5).}
	\label{fig: 126-130 analysis}
\end{figure}

In this section, we compare the differences in alpha-decay half lives
obtained using the various empirical equations 
discussed in Sec.~\ref{formulae} with respect to the average value for each nucleus.
The difference in the half lives, $\log_{10} T_{1/2}$ with respect to the average values
are shown in Fig.~\ref{fig: 114-118 analysis}, \ref{fig: 120-124 analysis} and \ref{fig: 126-130 analysis}
for $Z = 114 - 118$, $Z = 120 - 124$ and $Z = 126 - 130$, respectively.

It easily noticeable that the 
VS2 parameters set consistently predict half lives which are longer than
the average values for all isotopic series.
It fact, VS2 half lives are the longest out of the 8 empirical equations
for a given nucleus, except for some exceptions in the $Z = 126, 128$ and $130$ isotopes.
The PS parameters set also predicts 
a longer-than-average half lives, similar to the VS2,
although the differences from average values decrease (i.e. closer to average value) when going towards
heavier $Z$ numbers.
It is also interesting to note that the differences in half lives from average values
for both VS2 and PS are rather constant for all considered isotopic series.

On the other hand, the MB1 give the most fluctuation away from the average values across a given isotopic chain.
In Fig.~\ref{fig: 114-118 analysis}, the MB1 shows half lives which are shorter than the average values for $Z = 114$ to $118$.
For heavier elements shown in Fig.~\ref{fig: 120-124 analysis} and Fig.~\ref{fig: 126-130 analysis},
the pattern is reversed where the MB1 results are now longer than average values.
The MB2 parameters set, on the other hand, underestimate (i.e. giving shorter) the half lives 
as compared to the average values
but with almost constant pattern across a given isotopic chain.

We now move to compare similiarity in the patterns shown in
Fig.~\ref{fig: 114-118 analysis}, \ref{fig: 120-124 analysis} and \ref{fig: 126-130 analysis}
between different parameters sets.
The first group would be the original Brown parameters (denoted as B in the plots) and its modified versions MB1 and MB2.
We find that the trend for B and MB1 as a function of neutron number to be rather similar for all isotopic series.
On the other hand, results obtained with both B and MB1 appears to vary greatly from those using MB2.

\begin{figure}[hbt!]\centering
	\includegraphics[width=0.75\textwidth]{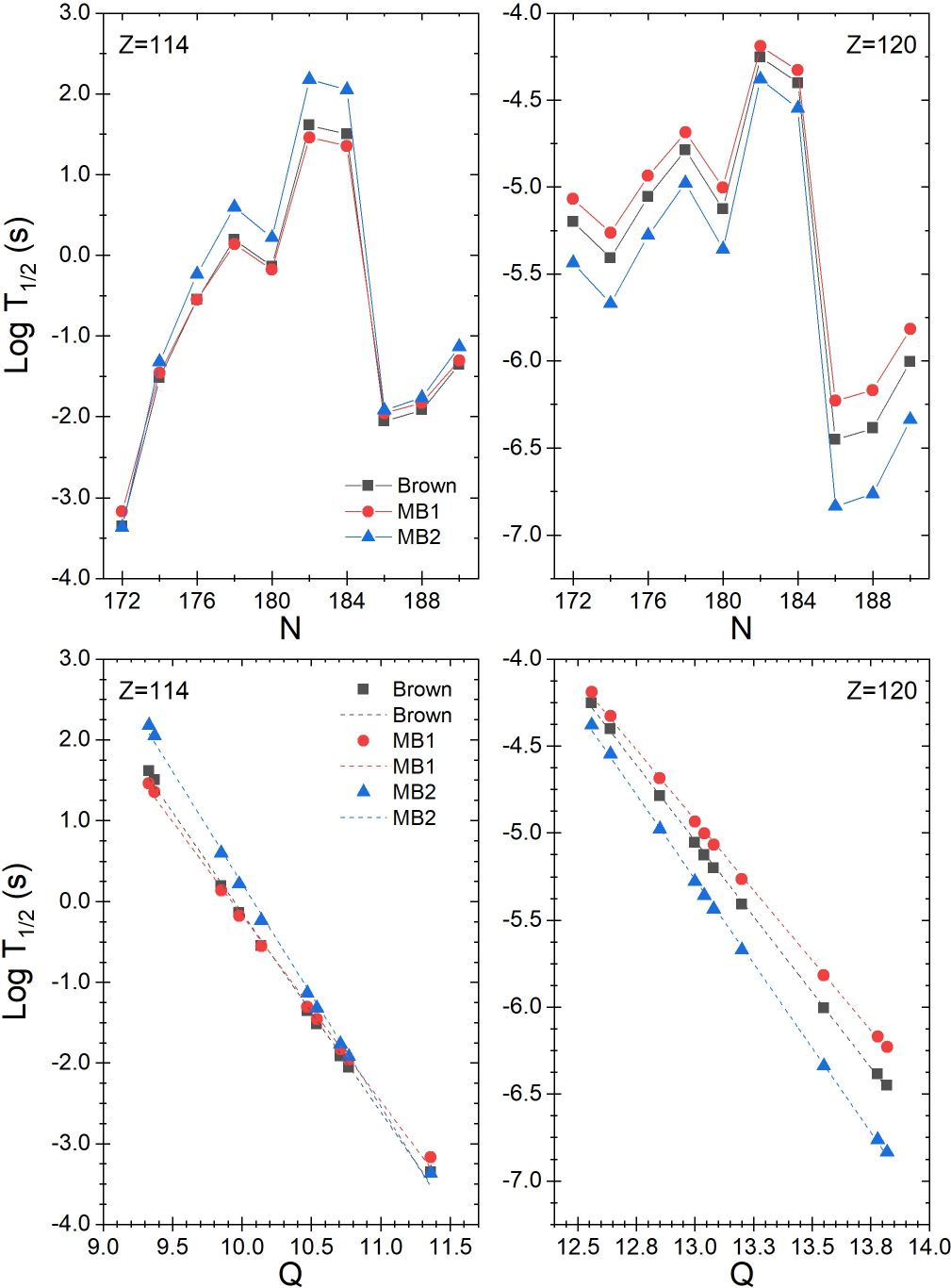}
	\caption{Analysis of Brown formulae type (B, MB1, and MB2) on $Z=114$ and $Z=120$.}
	\label{fig: half-lives vs Q alpha}
\end{figure}

The reason for such a huge variation in the pattern stems from the $Q_{\alpha}$
which will be addressed here.
Fig.~\ref{fig: half-lives vs Q alpha} shows the plot of the half lives 
as a function of neutron number, $N$, and $Q_{\alpha}$
for two isotopic samples of $Z = 114$ and $120$ obtained with the SkM* parametrization.
For $Z = 114$, we find that the half lives from B and MB1 to be extremely close.
Results from these two sets begin to differ from the MB2 set between
$N = 176$ and 184.
The $Q_{\alpha}$ within this neutron number region falls between $9.3$ to $10.2$ MeV 
(see bottom panel of Figure \ref{fig: half-lives vs Q alpha}).
At around this $Q_{\alpha}$ values, the MB2 half lives differs from the B and MB1 sets 
whereby the differences increases with decreasing $Q_{\alpha}$.
This explains the large variation in half lives between $N = 182$ and $184$ for MB2 as compared to B and MB1.

Similar dependence of half lives on $Q_{\alpha}$ to explain the huge difference between B and MB1 from the MB2
is also seen for $Z = 120$.
In this isotopic chain, the MB2 predicts shorter half lives as compared to B and MB1.
This is attributed to the $Q_{\alpha}$ values being larger than $12$ MeV.
Supposed that the $Q_{\alpha}$ values for these $Z = 120$ nuclei is lower than $12$ MeV,
the half lives calculated using MB2 will, on the other hand, be longer as compared to those of M and MB1 sets.

The next group of semi-empirical formula which are of the same origin are the R and MR parameters sets.
We found that the results obtained with these sets give rise to the same pattern with only slight variation between them.
This is to be expected since the two sets vary only  
by a slight change in the parameters entering the half-life equations.

To conclude this section, we also note that as one moves towards heavier elements, two distinct groups emerge.
On one hand, there are the MB2, VS1, R and MR parameters sets 
which predict shorter half lives as compared to the average values.
The other group made up of the MB1, VS2, B and PS parameters sets which predicts longer half lives.

\newpage

\newpage

\newpage

{\clearpage}
\newpage
\bibliographystyle{ws-ijmpe}
\bibliography{ref}

\end{document}